\documentclass[12pt]{article}
\pdfoutput=1 

\usepackage{setspace}
\usepackage[top=1in, bottom=1in, left=1in, right=1in]{geometry}
\usepackage{enumitem} 
\setlist{nolistsep} 

\usepackage{mathtools,amssymb,amsthm,amsthm,amsfonts,latexsym,bbm}
\usepackage{amscd,amsxtra} 
\usepackage{graphics,graphicx,xcolor}
\usepackage{ifpdf}
\usepackage[caption=false]{subfig}
\usepackage{multirow}
\usepackage[colorlinks,linkcolor=blue,citecolor=blue,urlcolor=blue]{hyperref}
\usepackage[comma]{natbib}
\usepackage{comment}
\usepackage{ulem}
\usepackage{authblk} 

\usepackage{hyperref}

\def\bSig\mathbf{\Sigma}

\def\bv{\boldsymbol b}

\def\ev{\boldsymbol e}

\def\uv{\boldsymbol u}

\def\xv{\boldsymbol x}
\def\yv{\boldsymbol y}
\def\zv{\boldsymbol z}

\def\Iv{\boldsymbol I}

\def\Lv{\boldsymbol L}

\def\Sv{\boldsymbol S}

\def\Vv{\boldsymbol V}

\def\Xv{\boldsymbol X}

\def\Zv{\boldsymbol Z}

\newcommand{\Ic}{\mathcal{I}}

\newcommand{\Lc}{\mathcal{L}}

\newcommand{\Uc}{\mathcal{U}}

\newcommand{\Wc}{\mathcal{W}}
\newcommand{\Xc}{\mathcal{X}}

\newcommand{\deltav}{\boldsymbol \delta}

\newcommand{\etav}{\boldsymbol \eta}

\newcommand{\muv}{\boldsymbol \mu}

\newcommand{\Sigmav}{\boldsymbol \Sigma}

\newcommand{\Omegav}{\boldsymbol \Omega}

\def\1v{\boldsymbol 1}
\def\0v{\boldsymbol 0}
\def\Id{\boldsymbol I} 

\newcommand{\Ind}[1]{\mathbbm{1}_{\left\{ {#1} \right\} }}

\newcommand{\Real}{\mathbb R}

\newcommand{\E}{\mathbb E}
\newcommand{\sgn}{\mathop{\mathrm{sign}}}
\def\Pr{\mathrm P}

\newcommand{\Cov}{\mathop{\rm Cov}}

\newcommand{\tr}{\mathop{\rm trace}}

\newcommand{\wh}{\widehat}

\newcommand{\argmax}{\operatornamewithlimits{argmax}}

\newcommand{\norm}[1]{\|#1\|}
\newcommand{\abs}[1]{\left\vert#1\right\vert}

\def\half{\frac{1}{2}}

\def\ie{\textit{i.e.}}

\theoremstyle{plain}
\newtheorem{theorem}{\sc Theorem}

\newtheorem{prop}[theorem]{\sc Proposition}

\theoremstyle{remark}

\theoremstyle{definition}

\newtheorem{assumption}{\sc Assumption}

\usepackage[figuresright]{rotating}

\title{A Statistical Approach to Set Classification by Feature Selection with Applications to Classification of Histopathology Images}	 
\author{Sungkyu Jung\thanks{Corresponding author}}
\affil{Department of Statistics\authorcr University of Pittsburgh\authorcr Pittsburgh, Pennsylvania 15260, U.S.A.\authorcr
E-mail: \texttt{sungkyu@pitt.edu}}
\author{Xingye Qiao}
\affil{Department of Mathematical Sciences\authorcr Binghamton University, State University of New York\authorcr Binghamton, New York 13902-6000, U.S.A.\authorcr E-mail: \texttt{qiao@math.binghamton.edu}}

\date{}

\begin{document}

\maketitle

\newpage
\begin{abstract}
\noindent Set classification problems arise when  classification tasks are based on sets of observations as opposed to individual observations. In set classification, a classification rule is trained with $N$ sets of observations, where each set is labeled with class information, and the prediction of a class label is performed also with a set of observations. Data sets for set classification appear, for example, in diagnostics of disease based on multiple cell nucleus images from a single tissue. Relevant statistical models for set classification are introduced, which motivate a set classification framework based on context-free feature extraction. By understanding a set of observations as an empirical distribution, we employ a data-driven method to choose those features which contain information on location and major variation. In particular, the method of principal component analysis is used to extract the features of major variation. Multidimensional scaling is used to represent features as vector-valued points on which conventional classifiers can be applied. The proposed set classification approaches achieve better classification results than competing methods in a number of simulated data examples. The benefits of our method are demonstrated in an analysis of histopathology images of cell nuclei related to liver cancer.
\end{abstract}

%

\vspace{0.15in} 

\newpage
\setlength{\belowdisplayskip}{0.5em} \setlength{\belowdisplayshortskip}{0.5em}
\setlength{\abovedisplayskip}{0.5em} \setlength{\abovedisplayshortskip}{0.5em}

\pagenumbering{arabic}
\setcounter{page}{1}
\section{Introduction}\label{sec:1intro}
\begin{figure}[!h]
\begin{center}
  {\subfloat[nuclei from normal tissue]
      {\label{fig:nuclei_img_normal}
      \includegraphics[width=.5\textwidth, trim = 20mm 100mm 20mm 110mm]{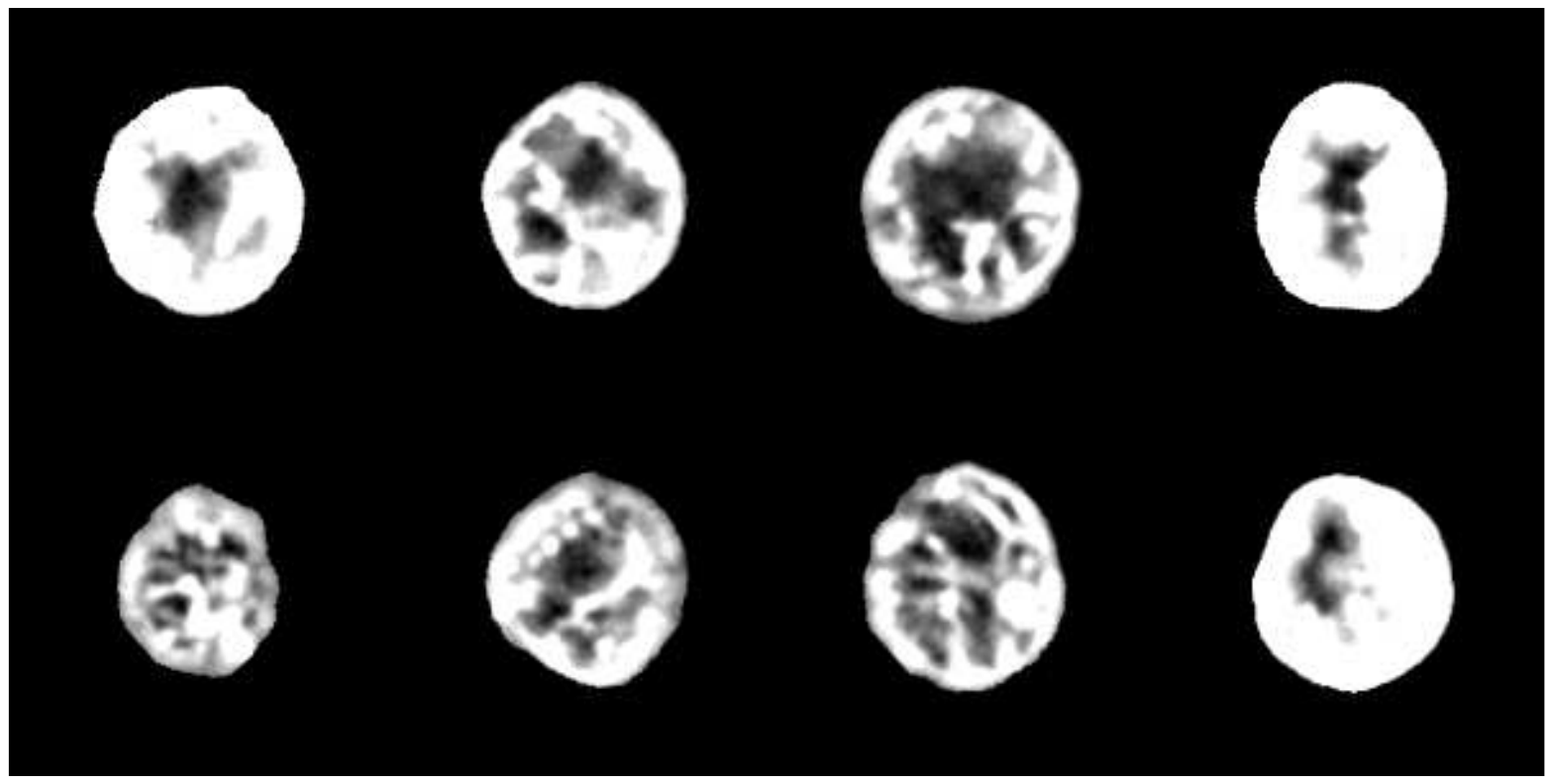}}
  \subfloat[nuclei from hepatoblastoma tissue]
      {\label{fig:nuclei_img_hepatoblastoma}
      \includegraphics[width=.5\textwidth, trim = 20mm 100mm 20mm 110mm]{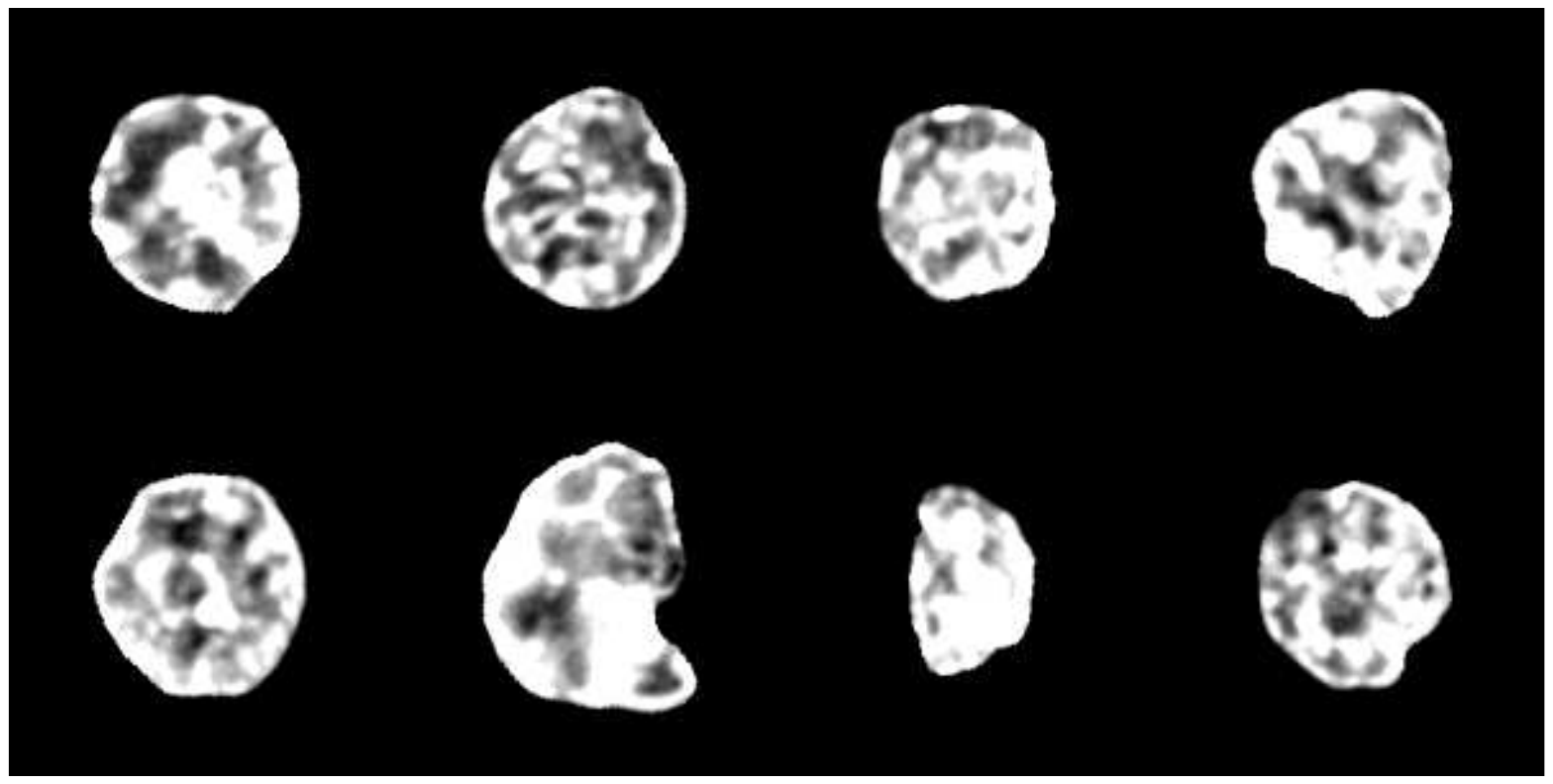}}
      }
\end{center}
\caption{Eight images of cell nuclei in a set labeled as normal tissue (a), and those from a set labeled as hepatoblastoma tissue (b). Each gray-scale image has $192 \times 192$  pixels. The set classification makes use of the set membership, to discriminate normal sets from hepatoblastoma sets.  \label{fig:nuclei_imags}}
\end{figure}

As advances in technology ease semi-automated segmentation and preprocessing of cell nucleus images, more pathologists are relying on  histopathology to discriminate diseased tissues from benign tissues. The classification of tissues based on microscopic examination is often achieved using many cell nucleus images.
Figure~\ref{fig:nuclei_imags} illustrates two sets of cell nucleus images from human liver tissues. The eight nuclei in the left panel  belong to a set labeled as normal tissues, while another set of eight nuclei from hepatoblastoma tissues is shown on the right. See Section~\ref{sec:realdata} for the background and analysis of the data.
An eminent statistical task to aid pathologists is to develop a method to classify a new tissue sample consisting of many nucleus images into the normal or the malignant.
The classification rule in need is learned from sets of observations and also should be able to predict a single class label for a new set of observations.
Such a problem, which we call \textit{set classification}, has not been studied much in the statistical literature, although it appears to be useful in image-based pathology \citep{Samsudin2010,Wang2010}.

To precisely define the set classification problem, suppose there are $N$ tissue samples, each of which is represented by a set $\Xc_i$ consisting of $n_i$ images of cell nuclei $\xv_{ij}$ in the tissue, \textit{i.e.}, $\Xc_i = \{\xv_{i1},\ldots, \xv_{in_i} \}$. Each set $\Xc_i$ or the corresponding tissue sample has its label $y_i$ (say, $y_i \in \{$normal, cancerous$\}$). Based on these $N$ sets of cell nuclei images, we wish to predict the label $y^\dagger$ for a new tissue sample containing ${n}_\dagger$ images, $\Xc_\dagger = \{\xv_{1}^\dagger,\ldots, \xv_{n_\dagger}^\dagger \}$.

A few characteristics of such a data set make the task challenging.
First, the order of observations $\{\xv_{i1},\dots,\xv_{in_i}\}$ in each set $\Xc_i$ may be only given by convenience, meaning that there is no correspondence between observations in different sets. In such a case,  $\xv_{i1}$ and $\xv_{\iota 1}$ in different sets ($i \neq \iota$) should not be directly compared. Moreover, the number of observations $n_i$ in each set may be different from one another.
These characteristics rule out a  naive approach for set classification, which is based on a long vector consisting of the $n_i$ observations $\Xv_i = [\xv_{i1}^T,\ldots,\xv_{in_i}^T]^T$. Instead, a more appropriate statistical model is obtained when a set is regarded as an empirical distribution of observations. 

\begin{figure}
\begin{center}
\includegraphics[width=.9\textwidth, trim = 20mm 110mm 20mm 110mm]{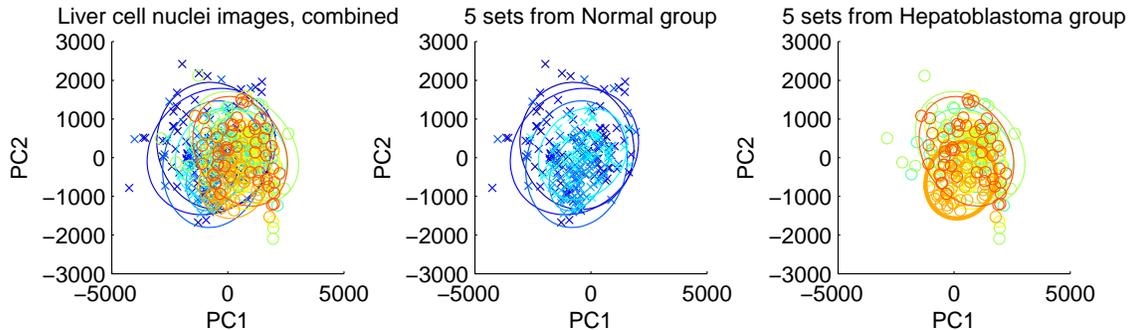}
\end{center}
\caption{Scatters of cell nucleus images with their mean and covariance as ellipses for each of 10 sets, projected onto the first two principal component directions. This figure appears in color in the electronic version of this article.\label{fig:nuclei_imgs_scatter}}
\end{figure}

To facilitate the understanding of the set as a distribution, a scatterplot of the liver cell nuclei data is overlaid with contours of the estimated normal densities in Figure~\ref{fig:nuclei_imgs_scatter}. A point in Figure~\ref{fig:nuclei_imgs_scatter} corresponds to an image of cell nucleus and each is from a tissue labeled as normal or hepatoblastoma  if marked by \textsf{x} or \textsf{o}, respectively.  Each  observation in a set is assumed to be drawn from a distribution, whose mean and covariance are represented by the contour of the density.
A useful insight of the data is that different sets have different distributional parameters. A visual inspection of the plot leads us to believe that the mean and covariance parameters will be useful for classification of sets. Classification based on the parameters of the distribution, or the features of the set, is the initial idea which we develop further in this paper.

Previous approaches to the classification of sets of images are mostly focused on extracting context-based features. For example,  \cite{Tsantis2009}  used morphological features and \cite{Wang2010}  used shape and texture features.  To the best knowledge of the authors, \cite{Ning2008} and \cite{Wang2010} were among the first to use set information in classification. Although not stated explicitly, they seem to have assumed a simple model of set classification, in that the set information of observations is not used in training but only  in prediction.

Having examined more general set classification models, we are proposing a context-free feature selection approach to set classification. Our method is based on extracting statistical features of sets such as the mean and principal components \citep[PCs;][]{Jolliffee2002}.
The proposed feature extraction--selection method transforms empirical PC directions and subspaces into feature vectors so that  conventional classifiers can be applied in the transformed space.
Multidimensional scaling \citep{Borg2005} is extensively used in mapping PC subspaces into a real vector space.
Our  procedure is not a single method but a general framework which can be coupled with any off-the-shelf classifier. We demonstrate the use of variants of linear and quadratic discriminant analysis \citep{Friedman1989,Srivastava2007}, Support Vector Machine \citep{Cristianini2000} and Distance-Weighted Discrimination \citep{Marron2007,Qiao2010Weighted} in this article. Prediction of the class label for a new observation is also based on its extracted features using  multidimensional scaling.

Since we use only statistical features which are free of context, the proposed approach can be applied to many other types of data beyond images.


In the next section, we formulate relevant statistical models for set classification. We then propose our set classification framework in Section~\ref{sec:3method} and feature selection procedure in Section~\ref{sec:4pcspaceselection}.
A simulation study is presented in Section~\ref{sec:numstudies} to examine the effectiveness of our methods in high-dimensional situations. In Section~\ref{sec:realdata}, the proposed methods are illustrated with a histopathological cell nucleus image data set. We conclude the article with a discussion in Section~\ref{sec:discussion}.

\section{Statistical models for set classification}\label{sec:2statisticalmodels}
In this section we describe statistical models from which the set classification problem arises.
 In set classification, each observation $\xv$ is labeled by two indices: $y \in \{1,\ldots, K\}$ for the class label and $i \in  \{1,\ldots, N\}$ for the set membership. The distribution of $\xv$ is different for different sets and classes. A natural model incorporating this characteristic of the data is a model with hierarchy, \textit{i.e.}, a top level for the class label and a lower level for the set  membership.

In the hierarchical model, a random observation $\xv$ follows a probability distribution $F(\xv \mid \theta)$ where $\theta$ is the idiosyncratic parameter of each set. The parameter $\theta$ of a set is considered as a random variable whose distribution is determined by a different class label, $y$. That is, the dependence of $\xv$ on the class label $y$ is  defined only  through its set membership.
Denote $\Theta_i$ as a random parameter of the distribution of the $i$th set. The hierarchical model for set classification is
\begin{align}
	 \Theta_i \mid (Y_i = k) &\sim H_k  \quad ( k = 1,\ldots, K), \label{eq:hierarchicalmodel}\\
	 \Xc_i \mid (\Theta_i = \theta) & \sim f(\Xc_i;\theta) = \prod_{j=1}^{n_i} f(\xv_{ij};\theta) \quad (i = 1,\ldots,N), \nonumber
\end{align}
where $H_k$ is the distribution of the parameter $\theta$ for the $k$th class. In the model above, we have assumed that the observations $\xv_{i1},\ldots,\xv_{in_i}$ in a set $\Xc_i$ are independent and follow an absolutely continuous distribution with  density function $f$.

To elucidate the use of the  hierarchical model, consider a model from which the data in Figure~\ref{fig:nuclei_imgs_scatter} may follow. Assuming normal distribution for each observation $\xv_{ij}$, it is visually evident from Figure~\ref{fig:nuclei_imgs_scatter} that the sets (or the corresponding  distributions) within either the normal or the hepatoblastoma group have different means and covariance matrices. The difference, however, is small compared to the difference between the normal and hepatoblastoma groups. The (random) parameters $\Theta_i = (\muv_i, \Sigmav_i)$ of the $i$th set may be modeled as $\muv_i | (Y = k)  \sim N_p(\delta_k, \sigma^2\Id_p)$ and $\Sigmav_i \mid (Y = k) \sim m^{-1}\Wc_p(\Sigma_{(k)}, m)$, where $\Wc_p(A,m)$ denotes the Wishart distribution of dimension $p$ with mean $A$ and $m$ degrees of freedom. The hyper-parameters $(\delta_k, \Sigma_{(k)})$
differ by the class label $k \in \{$normal, hepatoblastoma$\}$. The set $\Xc_i$ is then a set of i.i.d. observations following $N_p(\mu_i,\Sigma_i)$.

\begin{figure}
\begin{center}
      \includegraphics[width=0.8\textwidth, trim = 20mm 70mm 20mm 70mm]{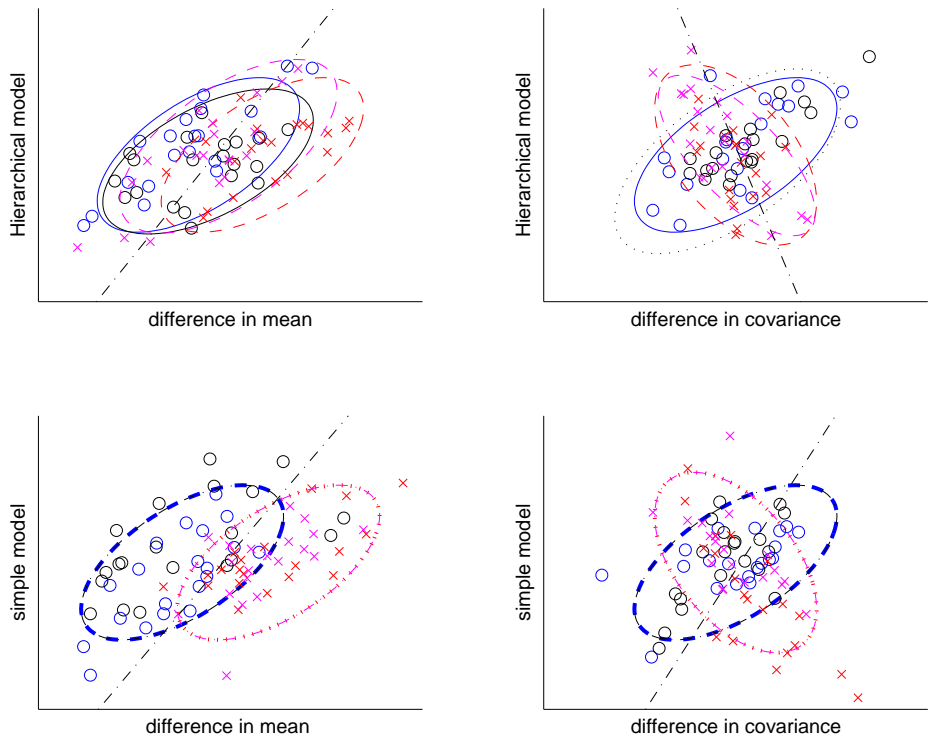}
\end{center}
\caption{Toy data examples from the hierarchical (top) and the simple (bottom) models. Hierarchical models allow different distributions (depicted as the shapes of ellipse) for different sets, while the simple model ignores the set membership. A linear separating hyperplain may not be useful as can be seen in the right panels. This figure appears in color in the electronic version of this article. \label{fig:toyexample}}
\end{figure}

Other examples of the hierarchical model include  situations where only the mean parameter $\muv_i$ is random while the covariance parameter $\Sigmav_i$ is fixed, or vice versa. These models are exemplified in the top two panels of Figure~\ref{fig:toyexample}.
Their population structures, displayed as ellipses, clearly visualize the underlying hierarchical model. In both examples, classification of sets can be done by utilizing the parameters (features) of the distribution, represented as the  locations and orientations of the ellipses. The separating hyperplane in Figure~\ref{fig:toyexample} by the linear discriminant analysis, pooling all observations in different sets together, is clearly less useful than using the distributional features.

A special case of the hierarchical model can be obtained by fixing the random $\Theta_i$ conditioned on $Y = k$ as a deterministic $\theta_k$. This model is then no longer hierarchical and is referred to as a simple model.
As illustrated in the bottom panels of  Figure~\ref{fig:toyexample}, all observations with the same class label have a common distribution.
Since set membership plays no role here, a conventional classification method with the weighted voting strategy is Bayes optimal (see Web Appendix A).
In our experience with simulated data analysis, our method based on the hierarchical model is superior to the weighted voting method, which is shown in Section~\ref{sec:numstudies}.



\section{Methods}\label{sec:3method}

Our framework for  set classification is two-step. First, features of a set, or the parameters of the corresponding probability distribution, are estimated, selected and transformed.
Then, a classifier is built based on the  features extracted from the training data set.

\subsection{Feature extraction}\label{sec:extraction}

In choosing appropriate features for sets,  it is worth revisiting  the characteristics of the liver cell nuclei data illustrated in Figure \ref{fig:nuclei_imags} and \ref{fig:nuclei_imgs_scatter}. First, each observation in the data lies in a fairly high-dimensional space while the number of sets and the number of observations are relatively small. Modeling with a large number of parameters is problematic due to collinearity and can lead to overfitting.
Second, as can be seen in Figure~\ref{fig:nuclei_imgs_scatter}, the location and orientational information of the distribution (the principal axis of the ellipses) is useful to visually separate sets between different groups.
These findings are what motivated us to consider using principal component analysis  \citep[PCA;][]{Jolliffee2002} to learn orientational information as well as the sample mean for the location to extract succinct but useful features of a set.

In the following, we discuss a method of extracting the orientation of the distribution through PCA. An application of multidimensional scaling (MDS) follows in order to represent the principal component features in a low-dimensional vector space. The vector-valued features which contain orientational information are then attached to the sample means for use in the sequel.

 \subsubsection{Principal component spaces}\label{sec:pcspaces}
Denote the sample mean of the observations in the $i$th set ($i = 1,\ldots,N$) by $\wh\muv_i = n_i^{-1}\sum_{j=1}^{n_i} \xv_{ij} \in \Real^p$. The empirical covariance matrix $\wh\Sigma_i \equiv n_i^{-1}\sum_{j = 1}^{n_i} (\xv_{ij} -\wh\muv_i ) (\xv_{ij} - \wh\muv_i )^T$ for the $i$th set is eigen-decomposed into $\wh\Sigma_i = \Sigma_{\ell = 1}^p \lambda_{\ell i} \ev_{\ell i} \ev_{\ell i}^T$, where the eigenvectors (PC directions) $\{\ev_{1 i},\dots, \ev_{pi}\}$ are orthonormal to each other and the eigenvalues (PC variances) $\lambda_{1i} \ge \cdots \ge \lambda_{pi} \ge 0$ are in descending order.
In the high-dimensional, low-sample size situation, \textit{i.e.}, $p \gg n_i$, we have $\lambda_{\ell i} = 0 $ for all $\ell \ge n_i$.

The direction of the major variation is represented by the first few empirical PC directions $\{\ev_{1 i},\dots, \ev_{r i}\}$ for some $r$. However, using the direction vectors as features can lead to an erroneous conclusion. Since the PC directions are axial, \textit{e.g.}, $\ev_{1 i} = \uv$ is essentially the same as $\ev_{1 i} = -\uv$, a special care is needed. Moreover, denote $(\ev_{\ell}^o, \lambda_{\ell}^o)$ for the eigen-pair from the population covariance matrix. The smaller $| \lambda_{1}^o - \lambda_{2}^o|$ is, the larger the variation in the empirical PC directions $\ev_{1 i}$ and $\ev_{2 i}$. In the worst case, where $\lambda_{1}^o = \lambda_{2}^o$, the population PC directions $\ev_{1}^o$ and $\ev_{2}^o$ are indistinguishable and thus unidentifiable. In such a case, the estimates $\ev_{1 i}$ and $\ev_{2 i}$ are only meaningful if they are understood as a whole. To be more precise,
\begin{prop}\label{thm:identifiability and consistency}
If $\lambda_{1}^o = \lambda_{2}^o > \lambda_{3}^o$ then the subspace spanned by the first two population PC directions
$\Lc^{o} = \mbox{span}( \ev_{1}^o,\ev_{2}^o)$
is identifiable, and $\Lc^{(2)}_i = \mbox{span}(\ev_{1 i}, \ev_{2 i})$ is a consistent estimator of $\Lc^{o}$.
\end{prop}
A proof of Proposition~\ref{thm:identifiability and consistency} can be found in Web Appendix A, in which the consistency of subspaces is formally defined.

On this end, denote the subspace spanned by the first $r$ empirical PC directions of the $i$th set by
$\Lc_i^{(r)} = \mbox{span}(\ev_{1i},\dots, \ev_{ri}),~1\leq i\leq N,$
 for $1\leq r\leq \min(p,n_i-1)$. Which $r$ to use is an important matter for model selection, and in Section~\ref{sec:4pcspaceselection} we propose a data-driven method to choose $r$. The arguments in the current section are applied to any given $r$.

Note that the collection $\{\Lc_i^{(r)}: i =1,\ldots,N\}$ does not lie in a Euclidean vector space, but in the Grassmannian manifold $G(r,p)$ \citep{Chikuse2003}. Therefore instead of analyzing elements in the curved manifold, $\Lc_i^{(r)}$s  are mapped onto a linear vector space. Such a mapping cannot be obtained by conventional dimension reduction methods such as PCA but can be achieved by multidimensional scaling. We first point out a distance function between $\Lc_i^{(r)}$s, on which the mapping  is  defined.

\subsubsection{Distances between principal component spaces}\label{sec:pcspaces_distance}
The distance between two subspaces of $\Real^p$ can be measured in terms of the canonical angles  formed by the subspaces \citep[\textit{cf}.][]{Stewart1990}.
As an intuitive example for the canonical angles, consider a simple case where both subspaces are of dimension 1. The two subspaces are each spanned by unit vectors $\xv_1$ and $\yv_1$. Then the distance between the subspaces is measured as the \textit{smallest angle} $\theta$ formed by the basis vectors, \textit{i.e.}, $\cos\theta = |\xv_1^T \yv_1|$, which is also called the canonical angle. For more general situations, canonical angles are computed as follows. Without loss of generality, let $\Lc_1$ and $\Lc_2$ be two subspaces with dimensions $s$ and $r (\le s)$ respectively, and $\Lv_1$ and $\Lv_2$ be matrices consisting of basis vectors of $\Lc_1$ and $\Lc_2$, respectively. In particular, the $p \times r$ orthonormal matrix $\Lv_{i} = [\ev_{1i},\dots, \ev_{ri}]$ is a basis matrix for the PC subspace $\Lc_i^{(r)}$. Let $\gamma_\ell$ be the  $\ell$th largest singular value of $\Lv_1^T \Lv_2$. Then the canonical angles between $\Lc_1$ and $\Lc_2$ are
 $\theta_\ell(\Lc_1,\Lc_2) = \arccos(\gamma_\ell)$ for $\ell = 1,\dots, r.$

%

 A distance between $\Lc_1$ and $\Lc_2$ is now defined using the canonical angles. We use a modified Euclidean sine metric  \citep[\textit{cf}.][]{Stewart1990}, defined by $\rho_s(\Lc_1,\Lc_2) = c ({\sum_{\ell=1}^r \sin^2(\theta_\ell) })^{\frac{1}{2}},$
where $c > 0 $ is a constant. Since $\theta_\ell \in (0,\pi/2]$, $\rho_s(\Lc_1,\Lc_2)$ is at most $c \sqrt{r}$. The role of the constant $c$ is to make the distance $\rho_s(\Lc_1,\Lc_2)$ commensurate with other features such as the mean. Specifically, when measuring pairwise distances among the PC spaces $\Lc_i^{(r)}$s, we choose $c$ to be the average of the empirical total variance in $\Lc_i^{(r)}$s, $c = \frac{1}{N}\sum_{i=1}^N\sum_{\ell=1}^r\lambda_{i\ell}$.
 This choice of $c$ leads to good classification results, which are reported in Web Appendix B.

%

 \subsubsection{Mapping $\Lc_{i}^{(r)}$ via multidimensional scaling}\label{sec:method-mds}
 Multidimensional scaling (MDS) maps $N$ objects with distances $\rho_s(\Lc_i^{(r)},\Lc_j^{(r)})$ into corresponding points $\zv_1, \ldots, \zv_N$ in an $m$-dimensional space $\Real^m$.
 The vector-valued configuration $\{\zv_1, \ldots, \zv_N\}$ is found by minimizing a loss function, which measures how well the original distances  $\rho_s(\Lc_i^{(r)},\Lc_j^{(r)})$ are approximated by  $\norm{\zv_i - \zv_j}_2$.
 MDS can be used to map data in a non-Euclidean space into a vector space so that conventional multivariate methods (based on the Euclidean geometry) can be used. In extracting features corresponding to the PC spaces $\Lc_i^{(r)}$, we use MDS to construct a mapping from the non-Euclidean objects $\Lc_i^{(r)} \in G(r,p)$ into vector-valued points. We use the classical multidimensional scaling (CMDS) since it gives an analytic solution requiring no numerical optimization. Detailed discussion on methods of MDS can be found in \citet{Borg2005} and \citet{Cox2008}.
 Note that PCA is not applicable for $\Lc_i^{(r)}$, which lies on a curved manifold $G(r,p)$.

 \textbf{Mapping training sample:} In CMDS, the  matrix of configuration $\Zv = [\zv_1, \ldots, \zv_N]$ is obtained from the $N \times N$ matrix of pairwise squared-distances $\Delta = (\rho_s^2(\Lc_i^{(r)},\Lc_j^{(r)}))_{i,j}$.
 Writing the matrix of pairwise squared Euclidean distances between columns of $\Zv$ as $D(\Zv) = (\norm{\zv_i - \zv_j}_2^2)_{ij}$, we minimize the loss function
  \begin{align}\label{eq:loss_cmds}
     L(\Zv) = \left\|-\frac{1}{2}C \left(D(\Zv) - \Delta \right)C \right\|_F^2,
  \end{align}
 where $C = \Id_ N- \frac{1}{N}\1v_N\1v_N^T$ is the centering matrix, and $\norm{A}_F$ is the Frobenius norm of matrix $A$. An analytic solution that minimizes (\ref{eq:loss_cmds}) is available, due to \cite{Gower1966}.
 \begin{theorem}\label{eq:cmds_solution}
  Let $B = -\frac{1}{2}C\Delta C$. From the eigen-decomposition of $B = Q \Lambda Q^T$, let $\Lambda_+$ be the diagonal matrix consisting of the positive eigenvalues and $Q_+$ the corresponding columns of $Q$. 
   Then $ \Zv = \Lambda_+^{\half}Q_+^T \in \Real^{m \times N}$ minimizes (\ref{eq:loss_cmds}).
 \end{theorem}
 The configuration $\Zv$ lies in $m$-dimensional real vector space, where $m < \min{(p,N)}$ is the number of positive eigenvalues of $B$.
 Note that $B_{\Zv} = \Zv^T \Zv$ is the $N \times N$ Gram matrix (consisting of the inner products of columns of $\Zv$), and that if all eigenvalues of $B$ are nonnegative, $B_{\Zv}$ coincides with $B$. 
 The coordinate matrix $\Zv$ is the mapped image in $\Real^m$ of the PC spaces $\Lc_i^{(r)}$, based on which a classifier can be trained. 

 \textbf{Mapping for prediction:} For a new observation $\Xc_\dagger$, we wish to extend the loss function (\ref{eq:loss_cmds}) in order to map the new PC space $\Lc_\dagger^{(r)}$ for classification.
 We include the new observation in the loss function while the training points $\Zv$ are fixed.
 Define the $(N+1) \times (N+1)$ symmetric matrix of pairwise squared-distances
  \begin{align*}
     \Delta_\dagger =
     \left(\begin{array}{cc}
      \Delta & \deltav_{12} \\
     \deltav_{21} & 0
     \end{array}\right),
  \end{align*}
 where $\deltav_{12} = (\deltav_{21})^T = (\rho^2_s(\Lc_i^{(r)}, \Lc_\dagger^{(r)}))_{i=1,\ldots,N}$.
 Let $\Zv_\dagger(\zv)$ be the $(m+1) \times (N+1)$ matrix of coordinates, where the first $N$ columns of $\Zv_\dagger(\zv)$ represent the training points and the last column $\zv$ represents the new point.  
 That is,
   \begin{align*}
      \Zv_\dagger(\zv) =
      \left(\begin{array}{c}
       \Zv      \\
       \0v_N^T
      \end{array}
      \zv \right),
   \end{align*}
 where $\Zv$ is the minimizer of (\ref{eq:loss_cmds}). While the training configuration $\Zv$ spans $m$ dimensions, the new point $\zv$ may not be found in span($\Zv$). Therefore the dimension of $\Zv_\dagger$ is increased by 1. 
  We generalize (\ref{eq:loss_cmds}) to minimize
    \begin{align}\label{eq:loss_cmds_new_obs}
       L_N(\zv) = \left\|-\frac{1}{2} P_N \left(D(\Zv_\dagger(\zv)) - \Delta_\dagger \right) P_N^T \right\|_F^2,
    \end{align}
 where $P_N = \Id_{N+1} - \frac{1}{N}\1v_{N+1}[\1v_N^T ; 0]^T$ is a projection matrix that works similarly to the centering matrix $C$ except that the mean of the first $N$ columns (\textit{i.e.}, the training points $\Zv$) is used for the centering.

 For the purpose of  classification, the last coordinate of the new $\zv \in \Real^{m+1}$ is not needed, because a classification rule learned from $\Zv$ only uses the first $m$ coordinates. An analytic solution for the first $m$ coordinates of $\zv$, hereafter denoted by $\zv_\dagger$, is also available. Define
 \begin{align*}
 B^\dagger = -\frac{1}{2}P_N \Delta_\dagger P_N^T =
   \left( \begin{array}{cc}
    B &  \bv_{12}\\
    \bv_{21} &  b_2
    \end{array}\right).
 \end{align*}
 It can be seen that $\bv_{12}$ is the vector of inner products $(\zv_i^T\zv_\dagger)_{i=1,\ldots,N}$, \textit{i.e.},
 $\bv_{12} = \Zv^T \zv_\dagger =  Q_+ \Lambda_+^{\half} \zv_\dagger.$
 We then have the coordinates of the new point
 \begin{align}\label{eq:cmds_solution_new}
     \zv_\dagger = \Lambda_+^{-\half}Q_+^T \bv_{12}.
 \end{align}

 The following theorem justifies the use of (\ref{eq:cmds_solution_new}) as the feature vector of the new observation $\Xc_\dagger$.

 \begin{theorem}\label{thm:mds_solution}
 If $\zv_\dagger^T \zv_\dagger \le b_2$, then
 \begin{align}\label{eq:inthm:mds_solution}
 \hat{\zv} = \left(\begin{array}{c}
       \zv_\dagger     \\
       \pm \sqrt{b_2 - \zv_\dagger^T \zv_\dagger}
      \end{array}\right)
 \end{align}
 are the local minima of $L_N(\zv)$ (\ref{eq:loss_cmds_new_obs}). If the equality holds, \ie, $\zv_\dagger^T \zv_\dagger = b_2$, then (\ref{eq:inthm:mds_solution}) is the global minimum of $L_N(\zv)$. Moreover, $\zv_\dagger^T \zv_\dagger \le b_2$ holds if $B^\dagger$ is nonnegative definite.
 \end{theorem}

 To the best of our knowledge, Theorem~\ref{thm:mds_solution} is the first attempt to obtain a closed form solution for mapping new observations onto coordinates of training configuration by CMDS. The loss function $L_N(\zv)$ is a fourth order polynomial of multiple arguments, thus finding a global minimum of $L_N$ is a challenging task, involving iterative numerical methods \citep[\textit{cf}.][]{Nie2006}. Theorem~\ref{thm:mds_solution} provides sufficient conditions for the existence of closed form local (or global) minima of $L_N$. A sufficient condition for $B^\dagger$ to be nonnegative definite is that the distance involved be Euclidean \citep[Ch. 19]{Borg2005}. Although our metric  $\rho_s$ is not Euclidean, $B^\dagger$ appears to be nonnegative definite in data analyses we have encountered, and the use of $\zv_\dagger$ has shown a good performance.

 \subsection{Proposed set classification procedure}\label{sec:method-two-step}
 The principal component features obtained from the previous section are now combined with the sample mean $\wh{\muv}_i$, on which a classification rule is trained. Specifically, our set classification rule is learned in two steps:
 \begin{description}
   \item[Step 1:] (Feature extraction)\\
   (a) For a given $r$, obtain the sample mean $\wh{\muv}_i$ and PC spaces $\Lc_i^{(r)}$ for each set (Section~\ref{sec:pcspaces}); \\
   (b)  Use CMDS to obtain the empirical PC space features $\zv_i$ (Sections~\ref{sec:pcspaces_distance} and \ref{sec:method-mds});\\
   (c) Collect combined features $\zv_i^{*} = [\wh{\muv}_i; \zv_i] \in \Real^{p+m}$ $(i = 1,\dots,N)$.
   \item[Step 2:] Build a classification rule with any off-the-shelf methods, with inputs $(\zv^*_i,y_i)_{i=1,\dots,N}$. 
   \end{description}

 Prediction of the class label of a new observation $\Xc_\dagger = \{\xv_{1}^\dagger, \dots,  \xv_{n_\dagger}^\dagger\}$ is also done in two steps. First obtain the features of the new set, 
 $\zv_{\dagger}^* =  [\wh{\muv}_\dagger;\zv_{\dagger}],$
 where $\wh{\muv}_\dagger$ is the mean vector of the observation in the set $\Xc_\dagger$ and $\zv_{\dagger}$ is from (\ref{eq:cmds_solution_new}). The classification rule trained in Step 2 above is applied to the input $\zv_{\dagger}^*$ to predict the label $\hat{y}_\dagger$ of the new set $\Xc_\dagger$.


 \section{Principal component space selection}\label{sec:4pcspaceselection}
 The choice of an appropriate dimension $r$ for the PC subspaces is a crucial step in feature extraction. We propose selecting an optimal dimension using training samples without making reference to the specific classification rules used in Step 2 above. A test to check whether one should choose $r = 0$, \textit{i.e.}, discarding the PC features, is also considered.


 Our strategy for selection of the dimension $r$ of the PC spaces $\Lc_i^{(r)}$ is to use a modified Hotelling's $T$-squared statistic defined for each $1\le r \le R$, where $R = \min\{n_1,\ldots,n_N,p\}$ is the greatest meaningful dimension we can choose. Here, denote the coordinate matrix of $\{\Lc_i^{(r)}\}$ for each $r$ by $\Zv^{(r)} = (\zv_i^{(r)})_{i=1}^N$, as defined in Theorem~\ref{eq:cmds_solution}.
 Let $G_1$ and $G_2$ be a partition of $\{1,\ldots, N\}$, where the set $G_1$ (or $G_2$) contains the indices corresponding to group $1$ (or 2). Let $N_k = | G_k |$, $k= 1,2$. Denote the mean difference by
 $\etav_r = \bar{\zv}_{G_1}^{(r)} - \bar{\zv}_{G_2}^{(r)}$, where $\bar{\zv}_{G_k}^{(r)} = \frac{1}{N_k}\sum_{i \in G_k} \zv_i^{(r)}$,
  the pooled covariance matrix by
 $S_r = \frac{1}{N}\sum_{k=1}^2 \sum_{i\in G_k} (\zv_i^{(r)} - \bar{\zv}_{G_k}^{(r)})(\zv_i^{(r)} - \bar{\zv}_{G_k}^{(r)})^T,$
 and the diagonal matrix consisting of the diagonal elements of $S_r$ by $D_r = \mbox{diag}(S_r)$.

 The modified Hotelling's $T^2$ statistic for comparing two multivariate populations is the sum of squared marginal $t$-statistics
$ T(r) = \etav_r' D_r^{-1}\etav_r, \ (r = 1,\ldots, R)$ \citep{Srivastava2008}.
 We choose an $r$ that gives the greatest value of $T(r)$, \textit{i.e.},
 \begin{align}\label{eq:featureselection}
 \hat{r} = \hat{r}(T) = \argmax_{r= 1,\ldots,R} T(r),
 \end{align}
 since the greater $T(r)$ is, the more separable the groups are. A rationale for the diagonal matrix $D_r$ in $T(r)$ comes from the use of the CMDS in Section~\ref{sec:method-mds}. The coordinates of the features $\zv_i^{(r)}$ are chosen so that they are in fact the principal axes; there is no correlation between coordinates of $\Zv^{(r)}$. Moreover, with a small $N$, the inversion of the matrix $S_r$ may cause numerical artifacts. Thus it is natural to exploit only the variances while discarding the covariances of $\zv_i^{(r)}$. Accordingly, $D_r$ is used instead of $S_r$.

%
%


 A permutation test for usefulness of the chosen PC features $\{\zv_i^{(\hat{r})}\}$ is now discussed. The features $\{\zv_i^{(\hat{r})}\}$ may not contain any discriminating information when, for example, the sub-populations differ only by location. This is translated into the null hypothesis:
 $$H_0: \zv_i^{(\hat{r})},\ i = 1,\ldots,N, \ \mbox{ are identically disributed.}$$
The test statistic $T(\hat{r})$ tends to be large when the alternative hypothesis (not $H_0$) is true.

Since the group labels are exchangeable under the null hypothesis, the sampling distribution of $T(\hat{r})$ under $H_0$ is obtained by permuting the class labels. Let $\{T(r,b)\}_{r=1}^R$ be the re-calculated statistics using permuted data (the $b$th random permutation).
The test then rejects $H_0$ if the test statistic $T(\hat{r})$ is larger than the $100(1-\alpha)$th percentile of the permuted statistics $T_b = \max_{1\le r \le R} T(r,b)$ for a level of test $\alpha$. In other words, the $p$-value of $T(\hat{r})$ is given by
$p_{T(\hat{r})} = \frac{1}{B}\sum_{b=1}^B 1_{\{T(\hat{r}) \le T_b\}}$
using $B$ random permutations.

 The dimension $\hat{r} (\ge 1)$ chosen by (\ref{eq:featureselection})
 is used for the PC features if $p_{T(\hat{r})} < \alpha$; otherwise we update $\hat{r} = 0$ so that the insignificant PC spaces are discarded.

 When the true underlying model has no covariance difference between groups, the proposed permutation test greatly enhances the performance of our classification rules, as demonstrated in Web Appendix D. 

\section{Numerical studies}\label{sec:numstudies}

\subsection{Competing methods}\label{sec:numstudies_methods}
Our feature selection--set classification methods are denoted by \textit{PCF-`classifier.'} For example, PCF-LDA denotes a set classification rule trained by linear discriminant analysis (LDA) on the extracted features, as that obtained in Sections~\ref{sec:3method} and \ref{sec:4pcspaceselection}. For the level of the permutation test, $\alpha = 0.05$ was used. Classifiers considered in numerical studies include LDA, quadratic discriminant analysis (QDA), and Support Vector Machines \citep[SVM,][]{Cristianini2000}, Distance-Weighted Discrimination \citep[DWD,][]{Marron2007}, and Minimum Distance Empirical Bayes rule \citep[MDEB,][]{Srivastava2007}.

Competing methods include voting classifiers, including two previous methods in \cite{Ning2008,Wang2010}. Consider binary classification with $Y = \pm 1$ and equal class sizes. Using LDA trained from all training samples, ignoring the set memberships, each $\xv_j$ in  a new set $\Xc =\{\xv_1,\ldots,\xv_n\}$ can be classified to
$\hat{y_j} = \sgn\{(\xv_j-\frac{\wh\mu_+ + \wh\mu_-}{2})^T\wh\Sigma^{-1}(\wh\mu_+-\wh\mu_-)\}$. \cite{Wang2010} proposed to use a set classifier $\hat{y} = \sum_{j=1}^n \hat{y_j}$ determined by a majority vote of the individual predictions of $\xv_j$, thus called an \textit{LDA-MV} classifier.
Weighted voting of the individual predictors, on the other hand, takes place when an \textit{LDA-WV} classifier $ \sgn\{\sum_{j=1}^n(\xv_j-\frac{\wh\mu_+ + \wh\mu_-}{2})^T\wh\Sigma^{-1}(\wh\mu_+-\wh\mu_-)\}$ is used.
 The LDA-WV classifier is an optimal classifier under the simple model (\textit{cf.} Section~\ref{sec:2statisticalmodels}). Classifiers QDA-(MV,WV), MDEB-(MV,WV), SVM-MV and DWD-MV are defined similar to LDA-(MV,WV) using the chosen classifiers. Precise definitions of these classifiers are provided in Web Appendix E. \cite{Ning2008} proposed using SVM-MV in classifications of sets.

In summary, the set classification methods we have considered are categorized into the following.
\begin{enumerate}
\item Proposed set classifiers: PCF-LDA, PCF-QDA, PCF-SVM, PCF-DWD and PCF-MDEB;
\item Existing methods: LDA-MV and SVM-MV;
\item Other competing methods: LDA-WV, QDA-(MV,WV), MDEB-(MV,WV) and DWD-MV.
\end{enumerate}
%
%


\subsection{Simulation models and results}\label{sec:numstudies_simmodels_results}

We use the following four hierarchical models for binary set classification. Denote $\ev_k \in \Real^p$ for
the vector whose components are all zeros except for a 1 in its $k$th.
We have $\xv_{ij} \sim N_p(\muv_i, \Sigmav_i)$ $(j = 1,\ldots,n_i)$, where
$
\muv_i \mid (Y = k)  \sim N_p(\deltav_{(k)}, 10^{-2}\Id_p),\ \deltav_{(1)} = \delta \ev_1, \deltav_{(2)} = \0v,
$
and
\begin{enumerate}
  \item No covariance difference:
  $ \Sigmav_i \equiv \Omegav_p(\rho)$;
  \item Wishart:
   $\Sigmav_i \mid (Y = k) \sim \frac{1}{m}\Wc_p(\Vv_k,m)$,  where
      $\Vv_k = \Omegav_p(\rho) + \sigma^2 \ev_k \ev_k'$ and $m = 10$;
  \item Inverse Wishart:
  $ \Sigmav_i \mid (Y = k) \sim  p\Ic\Wc_p(\Vv_k,p)$, where
      $\Vv_k  = \Omegav_p(\rho) + \sigma^2 \ev_k \ev_k'$;
  \item von Mises--Fisher: $
   \Sigmav_i  = \Omegav_p(\rho) + \sigma^2 \uv_i \uv_i'$, where $\uv_i \mid (Y = k) \sim \mbox{vMF}(\ev_k,\kappa),$ and $\kappa = 100.$
\end{enumerate}
We  use a class of covariance matrices $\Omegav_p(\rho)$ to consider highly correlated variables \citep[p.129]{Srivastava2007}. The $\Omegav_p(\rho)$ is a modified auto-regressive covariance matrix and is $\Omegav_p(\rho) = \left( \sigma_i \rho^{\abs{i-j}^{\frac{1}{7}}}\sigma_j \right)_{ij}$, for $0 \le \rho \le 1$, where $\sigma_i$ are independently drawn from the uniform distribution on $[4/5, 6/5]$.
The von Mises--Fisher distribution used in model (4) is a Gausssian-like distribution for direction vectors, with mode at $\ev_k$ and a concentration parameter $\kappa$  \citep[\textit{cf}.][]{Mardia2000}.  In all of the models, $\muv_i$ and $\Sigmav_i$ are independent.

For each model, the performances of the methods in Section~\ref{sec:numstudies_methods} are evaluated by empirical misclassification rates.
The parameters of the models are set as $\delta = 1$, $\sigma = 3$, and $\rho \in \{0, .5\}$. We  tested the number of sets $N = 10, 20$ (equal class sizes) with $p = 20, 200, 400$, and the set size $n_i$ was independently selected by $n_i \sim \max \{ \lfloor N(20, 5^2)\rfloor, 10 \}$.

Table~\ref{tab:simulationResultsModel} summarizes the result of experiments based on 100 repetitions for $(p,N) = (20,10), (400,20)$. Throughout different models and various settings of dimension--sample size pairs, the proposed methods exhibit much smaller misclassification rates than the competing methods. This is not surprising because the models (2--4) from which the data are generated are the hierarchical models with a difference in PC structures. Since our method choose the PC spaces as features, good performance is to be expected.
Moreover, even when there is no covariance difference in model (1), the proposed methods are comparable to the weighted voting methods. In model (1) with $\rho = 0$, the permutation test in Section~\ref{sec:4pcspaceselection} makes the PC space features used only 2--8\% of the time, thus effectively discarding unimportant PC information.
On the other hand, in models (2--4) where the PC spaces possess discriminating information related to class, the tests resulted in rejecting the null hypothesis with empirical power of 95--100\%, thus successfully utilizing the PC space features.
When the variables are highly correlated ($\rho = 0.5$), our methods  show rates smaller or comparable to those obtained using MDEB-WV, which is designed to work well for correlated variables. The simulation result also confirms that LDA-WV is not optimal in the general hierarchical model.

\begin{sidewaystable}
\caption{Empirical misclassification rates (in \%, mean and standard error based on 100 simulations) illustrate that our methods PCF-`classifier' show better performance than  competing methods. Full tables including the results of all dimension-sample size pairs and all methods listed in Section~\ref{sec:numstudies_methods} are available at Appendix Table F.1 and F.2.}\label{tab:simulationResultsModel}
\begin{tabular}{cc|cccc|ccc}
\hline
         &           &   \multicolumn{4}{c}{Proposed methods (\textit{PCF-'classifier'})} & \multicolumn{3}{c}{Competing methods} \\  \hline
{Model}  & {$(p,N)$} &  {PCF-LDA}  & {PCF-SVM} &  {PCF-DWD} &{PCF-MDEB} &  {LDA-WV}& {SVM-MV}& {MDEB-WV} \\
\hline
\multirow{2}{*}{(1), $\rho = 0$}
   &  (20,10)  &25.41(5.43) & 22.03(5.45) & 18.75(4.28) & 18.26(4.42) & \textbf{18.18(4.60)} & 20.72(4.56) & 19.57(6.08) \\
   & (400,20) &33.18(3.86) & 33.22(3.89) & 32.61(3.97) & \textbf{32.59(3.93)} & 37.61(4.47) & 38.27(4.70) & 33.42(4.01) \\
\hline
 \multirow{2}{*}{(2), $\rho = 0$}
    &  (20,10) &\textbf{4.06(3.58)} & 5.20(4.39) & 6.56(4.76) & 6.14(4.60) & 38.18(4.72) & 39.95(4.80) & 34.34(7.26) \\
    &(400,20) &2.22(1.25) & 2.22(1.25) & \textbf{2.22(1.24)} & 2.22(1.26) & 47.80(3.66) & 46.55(3.37) & 43.31(3.31) \\
\hline
 \multirow{2}{*}{(3), $\rho = 0$}
    &  (20,10) &5.80(4.07) & \textbf{5.75(3.83)}   & 6.10(3.79)  & 7.34(5.69) & 36.10(6.33) & 36.74(6.14) & 34.81(12.20) \\
    &(400,20) &21.03(6.94) & 21.03(6.94) & \textbf{20.88(6.74)}   & 23.35(7.18) & 34.67(3.48) & 32.57(3.66) & 30.14(6.78) \\
\hline
\multirow{2}{*}{(4), $\rho = 0$}
    &  (20,10) & \textbf{13.78(5.66)} & 15.53(5.41) & 15.73(4.81) & 15.36(4.55)& 37.25(4.36) & 38.28(4.20) & 33.83(6.30) \\
    &(400,20) & 32.49(3.58) & 32.50(3.55) & 31.83(3.56) & \textbf{31.65(3.53)} & 38.24(4.94) & 38.72(4.39) & 33.55(3.92) \\
\hline
\multirow{2}{*}{(1), $\rho = .5$}
    &  (20,10) &23.69(5.49) & 20.57(4.96) & 17.44(3.96) & 17.09(3.99) & \textbf{16.56(4.07)} & 18.39(3.76) & 17.31(5.24) \\
    &(400,20) &33.93(4.75) & 34.01(4.81) & 34.28(4.76) & 34.41(4.82) & 36.13(4.51) & 36.08(4.12) & \textbf{31.39(3.96)} \\        				 \hline
\multirow{2}{*}{(2), $\rho = .5$}
    &  (20,10)&\textbf{3.91(4.23)}     &           4.68(4.55) & 6.16(4.95) & 5.71(4.81) & 39.52(4.66) & 40.25(4.94) & 34.06(7.11) \\
    &(400,20)&1.58(1.05) & 1.58(1.05) & \textbf{1.56(1.05)} & 1.58(1.05) & 48.24(3.46) & 47.00(3.31) & 43.45(3.35) \\
\hline
\multirow{2}{*}{(3), $\rho = .5$}
    &  (20,10) &5.55(4.46) & \textbf{5.46(4.14)}    & 5.96(4.26)     & 7.38(7.21) & 37.28(6.41) & 37.96(6.37) & 34.66(11.99) \\
    &(400,20) &20.40(7.09) & 20.46(7.13) & \textbf{20.19(6.96)} & 22.38(6.07) & 35.19(3.54) & 32.98(3.81) & 30.16(6.70) \\
\hline
\multirow{2}{*}{(4), $\rho = .5$}
    &  (20,10) &\textbf{1.97(1.70)} & 2.86(2.10) & 4.46(   2.77) & 3.76(2.29)  & 38.01(3.96) & 38.46(4.31) & 32.34(6.18) \\
    &(400,20) &33.32(4.42) & 33.34(4.40) & 33.89(4.88) & 33.95(5.01) & 37.83(4.73) & 37.48(4.40) & \textbf{31.80(3.52)} \\
 \hline
\end{tabular}
\end{sidewaystable}

  \section{Classification of liver cell nucleus images}\label{sec:realdata}\label{sec:realdata_liver}
  One of the common procedures used to diagnose hepatoblastoma (a rare malignant liver cancer) is biopsy--a sample tissue of a tumor is removed and examined under a microscope. A tissue sample contains a number of nuclei, a subset of which  is  then processed to obtain segmented images of nuclei. The data we analyzed contain 5 sets of nuclei from normal liver tissues and 5 sets of nuclei from cancerous tissues; see Figure~\ref{fig:nuclei_imags}. Each set contains $50$ images. The data set is available at {http://www.andrew.cmu.edu/user/gustavor/software.html}
  and is introduced in \citet{Wang2010,Wang2011}, in which one can find details on registration and segmentation of the nucleus images.

  We  tested the proposed method on the liver cell nuclei image data set in two different ways. First, using all 50 observations in each set, we evaluated the leave-one-out classification error rates by fixing one set as test data and training with the remaining 9 sets. This experiment was repeated for dimension-reduced data (using PCA for combined data). Table~\ref{tab:liver} summarizes the results. Our methods only misclassified the 9th set, showing the best performance (together with LDA-MV) among all methods considered. The 9th set was misclassified by many methods in different dimensions. This can be explained by a visual inspection of the scatterplot in Figure~\ref{fig:nuclei_imgs_scatter}. The orientation of the distribution of the 9th set, illustrated as the thicker orange ellipse in the right panel, is pointing north-east, which is closer to the orientation of the normal group than the north-west orientation of the hepatoblastoma group.

\begin{table}
  \caption{Misclassification rates and labels of misclassified sets in leave-one-out classification experiments of liver data. From original data with dimension $d = 22,500$ to dimension-reduced data by PCA. Results from QDA-(MV, WV) and DWD-MV are omitted as their performances are poor. \label{tab:liver}}
  \begin{center}
\begin{footnotesize}
  \begin{tabular}{c|cccc}
           & $d = 36,864$ &  $d = 100$ &  $d = 10$ & $d = 2$  \\
             \hline
  PCF-LDA  &   $1/10$,  (9th)&  $1/10$,  (9th) &    $0/10$   &    $0/10$  \\
  PCF-QDA  &   $1/10$,  (9th)&  $1/10$,  (9th) &    $0/10$   &    $2/10$,  (4,5)th  \\
  PCF-SVM  &   $1/10$,  (9th)&  $1/10$,  (9th) &    $0/10$   &    $0/10$  \\
  PCF-DWD  &   $1/10$,  (9th)&  $1/10$,  (9th) &    $0/10$   &    $0/10$  \\
  PCF-MDEB  &   $1/10$,  (9th)&  $1/10$,  (9th) &    $0/10$   &    $0/10$  \\
  LDA-MV   &  $1/10$,  (9th)&  $1/10$,  (9th) & $1/10$,  (9th)&  $1/10$,  (9th) \\
  LDA-WV &     $3/10$,  (3,7,9)th &  $0/10$   & $1/10$,  (9th)&  $1/10$,  (9th) \\
  SVM-MV &    $1/10$,  (9th)&   $2/10$, (6,9)th &    $1/10$,  (9th)&  $1/10$,  (9th) \\
  MDEB-MV &   $1/10$,  (9th)&  $1/10$,  (9th) &   $1/10$,  (9th)   &    $3/10$,  (3,7,9)th \\
  MDEB-WV &   $1/10$,  (9th)&  $1/10$,  (9th) &    $0/10$   &    $2/10$,  (7,9)th  \\

  \end{tabular}
\end{footnotesize}
  \end{center}
\end{table}

\begin{figure}
\begin{center}
\includegraphics[width=1\textwidth, trim = 10mm 70mm 10mm 70mm]{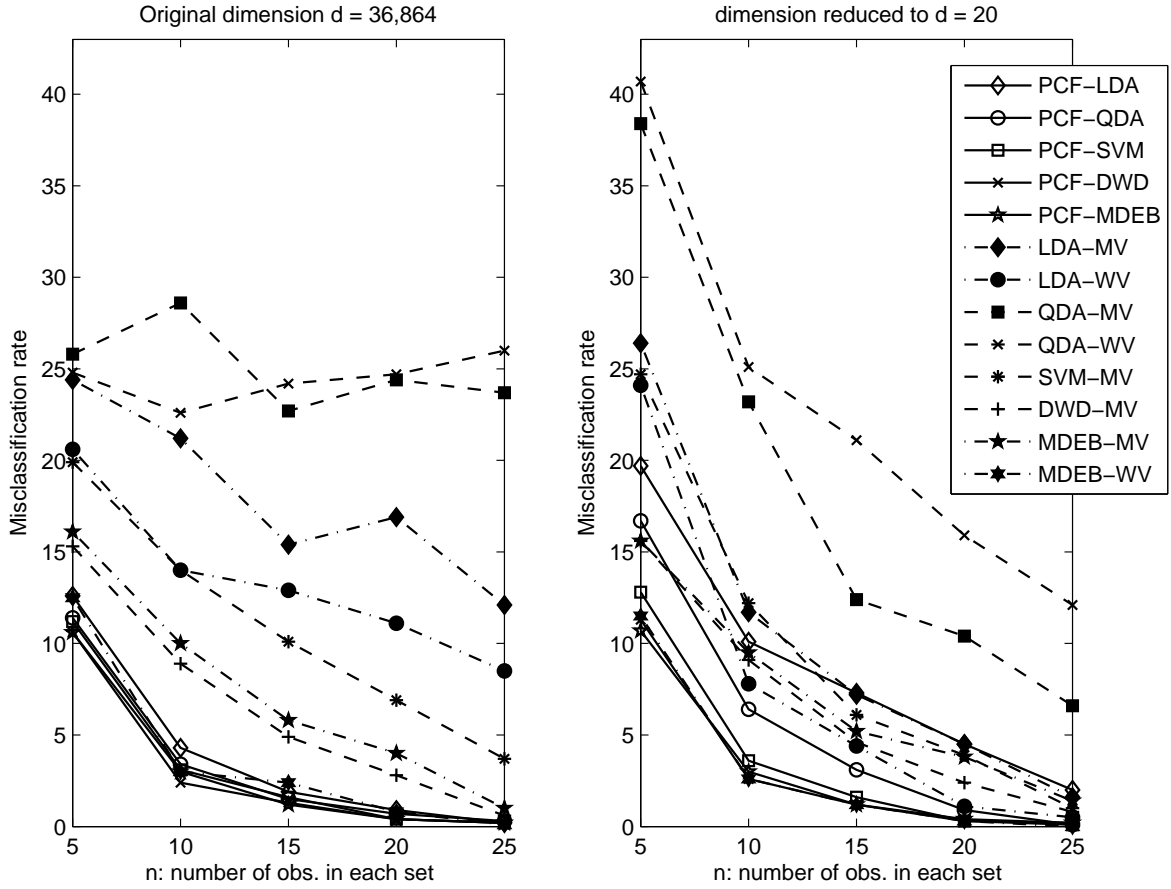}
\vskip -1in
\end{center}
\caption{Misclassification rates by different set size $n$ for the binary set classification of liver cell nuclei images. The proposed set classifiers exhibit much smaller misclassification rates than other methods when both $n$ and $N$ (the number of sets) are small. The weighted voting classifier with MDEB rule (MDEB-WV) shows comparable performance. An initial dimension reduction (PCA for combined data) improves overall performances of most classification methods. \label{fig:img_result} }
\end{figure}

Second,  the performances of the set classifiers were evaluated when the number of observations $n$ in each set is smaller than 50. Although technology enables processing of an image through semi-automatic registration and segmentation, the cost of obtaining images is still expensive. Thus a method exhibiting solid performance in small $n$ setting has a clear advantage over other methods. In particular, we randomly chose $n \in \{5, 10, 15, 20, 25\}$ images in each of $N = 10$ sets as training data and also chose non-overlapping $n$ images in each set as testing data.
After training the set classifiers, the empirical misclassification rate was computed using the remaining testing data. This procedure was repeated for 100 random subsets of size $n$, for each choice of $n$. Figure~\ref{fig:img_result} summarizes the results of the small-$n$ experiment. Our methods exhibited significantly better performance than the others, with the exception of MDEB-WV, for both the original and dimension-reduced data sets.

%
%

The features of the data we chose only include `statistical features' of the raw images, but do not include any other features that needs expert input, such as shape or texture features \citep[\textit{cf}.][]{Chen2006}.

\section{Discussion} \label{sec:discussion}
We have introduced a novel set classification framework and proposed a feature extraction--selection procedure. The method is based on the hierarchical modeling of set classification data. The features of sets, chosen by a data-driven approach, are used as inputs for any off-the-shelf classification method. When the orientation of major variations in sets possesses discriminating information, the proposed method adaptively uses such information for classification. Classical multidimensional scaling  with the modified Euclidean sine metric between PC subspaces are shown to work well in mapping nonlinear features  into vector-valued points. An analytic solution for the mapping of new observations is derived.

Note that \citet{Cox1993} and \citet{Witten2011} also proposed using multidimensional scaling (MDS) for classification. However, the method proposed in Cox and Ferry only works for vector inputs and uses a linear regression for mapping of new observations. Witten and Tibshirani proposed modifying the nonmetric multidimensional scaling, which requires an iterative algorithm to solve, but they did not consider using MDS for set classification.


There is the potential to extend our framework to more general cases. In particular, the independence assumption in the hierarchical model may be relaxed to model  dependencies among observations in a set. This is especially relevant in classification of tissues based on sets of images because the cells in a tissue have location information. Images observed with close locations can be assumed to have high correlation. We conjecture that including the location in the hierarchical model greatly increases the accuracy of classification. We have focused on the independent hierarchical model in this paper, as it serves well as an introduction to set classification.


The performance of our set classifiers can be greatly limited when the set sizes are severely imbalanced. The principal component estimates for each set use only $n_i$ samples. If $n_i$ vary greatly, so does the credibility of the estimated features. Moreover, the choice of the dimension $r$ for the PC features in Section~\ref{sec:4pcspaceselection} is also limited up to $\min(n_i)$. These problems are mitigated by excluding those sets with a small sample size or by applying a sensible dimension reduction, but the classifiers could be further honed to properly accommodate imbalanced set sizes.

Another issue is the use of classifiers. We have demonstrated the use of  conventional classifiers (LDA, QDA and linear SVM), but have not examined using more sophisticated methods such as kernel machines \citep[\textit{cf}.][]{Cristianini2000}. While use of kernel machines may benefit the classification task as shown in \cite{Wang2010}, we chose to focus on our feature selection procedure  using simpler methods. On the other hand, since our method is a general framework that works with any classifier, the presented methods have the potential to be improved by advanced kernel machines.

\section{Supplementary Materials}
Web Appendices referenced in Sections~\ref{sec:2statisticalmodels}, \ref{sec:extraction}, \ref{sec:4pcspaceselection} and \ref{sec:numstudies} as well as a zipped code and example data are available.

\section*{Acknowledgements}
We thank Gustavo Rohde for discussions on related topics and providing the histopathology data sets, and the Editor, Associate Editor and anonymous referees for their valuable comments. Jung was partially supported by NSF grant DMS-1307178. Qiao was partially supported by the Simons Foundation \#246649.

%
\bibliographystyle{asa}
\bibliography{scfs_bibfile}

\newpage
\begin{center}\Large
Web-based Supplementary Materials: ``A Statistical Approach to Set Classification by Feature Selection with Applications to Classification of Histopathology images''
\end{center}
\begin{center}\large
by Sungkyu Jung and Xingye Qiao\\University of Pittsburgh and State University of New York, Binghamton
\end{center}

\appendix
\section*{Summary}
This supplementary document for `A Statistical Approach to Set Classification by Feature Selection with Applications to Classification of Histopathology images' includes
\begin{enumerate}
  \item Optimal decision rules in set classification as referenced in Section~2 (Web Appendix \ref{sec:optimalclassifier}.1--\ref{sec:optimalclassifier}.3);
  \item A proof of Proposition~1 as referenced in Section 3.1.1 (Web Appendix \ref{sec:Proof of Proposition 1});
  \item A proof of Theorem~2 as referenced in Section 3.1.3 (Web Appendix \ref{sec:Proof of Theorem 2});
  \item A discussion on the scale factor as referenced in Section 3.1.2 (Web Appendix \ref{sec:scale factor});
  \item Additional tables and figures as referenced in Section 4 (Web Appendix \ref{sec:Comparison of different statistics} and \ref{sec:Validatio});
  \item Additional definitions and tables as referenced in Section 5 (Web Appendix~\ref{sec:LDAdefs} and \ref{sec:numstudiesTables}).
  \item A collection of Matlab codes for the proposed set classification and example data are available as a zipped file. 
\end{enumerate}

\section{Technical details}\label{sec:optimalclassifier}
In this section, we first provide theoretical optimal classifiers for the set classification models introduced in Section 2, and discuss their implications to the proposed method. In sections~\ref{sec:Proof of Proposition 1} and \ref{sec:Proof of Theorem 2} proofs of Proposition 1 and Theorem 3 are provided.

\subsection{Simple model}\label{sec:simple model}
The simple model for set classification, referenced in Section 2, is a special case of the hierarchical model and can be written as
\begin{align*}
\Xc_i \mid (Y = k) \sim \prod_{j=1}^{n_i}f_k(\xv_{ij})
\end{align*}
with $\Pr(\{Y = k\}) = \pi_k$.

Let $(\Xc, Y)$ be a new random object from the simple model.
Let $\Phi$ be a collection of set classification rules, and $\phi(\Xc) \equiv \phi(\{\xv_1,\ldots,\xv_n\}) \in \Phi$. The theoretical Bayes optimal decision rule $\phi^*$ minimizes the risk function  $R(\phi) =  \Pr(\phi(\Xc) \neq Y)$ among all $\phi \in \Phi$. Since
\begin{align*}
1 - R(\phi) & = \sum_{k=1}^K \pi_k\Pr(\phi(\Xc) = k \mid Y = k)\\
            & = \sum_k \pi_k \int \Ind{\phi(\Xc) = k} f_k(\Xc) d\Xc,
\end{align*}
the optimal $\phi^*$ must satisfy
$\phi^*(\Xc) = \argmax_k f_k(\Xc) \pi_k = \argmax_k \prod_{j=1}^n {f}_k(\xv_j) \pi_k$ for any $\Xc$.

With an assumption that each $f_k$ is a $p$-variate normal distribution with mean $\muv_k$ and covariance matrix $\Sigma$, a classifier using the linear discriminant analysis can be shown to be optimal in the following sense. For simplicity, consider binary classification with $Y = \pm 1$ and assume $\pi_{+}=\pi_{-}$. Then the Bayes decision rule is
\begin{align}
\phi^*&(\Xc)  = \argmax_{k\in\{+1,-1\}} \prod_{j=1}^n {f}_k(\xv_j)  \nonumber \\
& =\sgn\left\{\left(\prod_{j=1}^n {f}_+(\xv_j)/\prod_{j=1}^n {f}_-(\xv_j)\right)-1\right\}  \nonumber\\
& = \sgn\left\{\sum_{j=1}^n\left(\xv_j-\frac{\mu_++\mu_-}{2}\right)^T\Sigma^{-1}(\mu_+-\mu_1)\right\}. \label{eq:BayesOptimalSimple}
\end{align}
An estimator $\wh{\phi}^*(\Xc)$ of the optimal rule can be obtained by plugging-in the estimates $\wh\mu_+, \wh\mu_-$ and $\wh{\Sigma}$.
If the order of the sign operator and summation in  (\ref{eq:BayesOptimalSimple}) is switched, then a related, but suboptimal, classifier is obtained. The corresponding empirical classifier
\begin{align}\label{eq:LDA-mv}
\sum_{j=1}^n\sgn\left\{\left(\xv_j-\frac{\wh\mu_+ + \wh\mu_-}{2}\right)^T\wh\Sigma^{-1}(\wh\mu_+-\wh\mu_1)\right\}
\end{align}
works as if the set label is determined by a majority vote of the individual predictions of $\xv_j$. We call (\ref{eq:LDA-mv}) as
an \textit{LDA-MV} classifier. In contrast, the optimal classifier (\ref{eq:BayesOptimalSimple}) works as if the individual predictions vote with weights, thus called a \textit{LDA-WV} (weighted voting) classifier.

The LDA-WV is optimal only if the simple model can be assumed. In the hierarchical model, both LDA-MV and LDA-WV exhibit poor performances in the simulated and real data analyses, shown in Section 5 and 6. The  majority voting strategy has been suggested in \citet{Ning2008} and \citet{Wang2010}, but they have not discussed its relationship to the theoretical Bayes rule.

\subsection{Hierarchical model}\label{sec:Hierarchical model}
The hierarchical model (equation 1, Section 2) enables us to model appropriate characteristics of  set classification data. In the hierarchical model, the optimal classification rule for prediction of a new set $\Xc$ minimizes the risk $P(\phi(\Xc) \neq Y)$. Writing $\pi_k = P(\{Y = k\})$, we have
\begin{align}
	\Pr&(\phi(\Xc)=Y) =  \sum_{k=1}^K \pi_k\Pr(\phi(\Xc)=k\mid Y=k) \nonumber\\
       = & \sum_{k=1}^K \pi_k \E_{\Theta} \left[\Pr(\phi(\Xc)=k \mid \Theta) \mid Y = k\right]  \nonumber\\
	 = & \sum_{k=1}^K \int \Ind{\phi(\Xc)=k}\left(\pi_k\int f(\Xc;\theta) h_k(\theta)d\theta \right)d\Xc \label{eq:hierarchicalRisk2},
\end{align}
where $h_k(\theta)$ is the conditional density function of $\Theta$ given $Y = k$, provided that the density exists. From (\ref{eq:hierarchicalRisk2}), the optimal rule $\phi(\Xc)$ predicts $\hat{Y} = {k}$ if ${k}$ maximizes
\begin{align}\label{eq:hierrisk}
	\pi_k\int f(\Xc;\theta) h_k(\theta)d\theta.
\end{align}

Since the estimation of (\ref{eq:hierrisk}) is challenging, we look at the problem at a different angle. First consider the special case that the parameter $\Theta$ is fixed for each class $k$, \textit{i.e.}, $\Pr(\Theta = \theta \mid Y = k) = \Ind{\theta = \theta_k }$. Then  (\ref{eq:hierrisk}) is simplified to $ \pi_k f(\Xc ; \theta_k)$, which is the expression used in the optimal classifier of the simple model.
On the other hand, if the distribution of $\Xc$ given $\theta$ is concentrated at only one point, \textit{i.e.}, $ \Pr(\Xc = X \mid \Theta = \theta) = \Ind{g(X) = \theta}$ for an invertible function $g$, then the optimal classifier is
\begin{align}\label{eq:hierrisk2opt}
\phi(\Xc) = \argmax_{k = 1,\ldots,K} \pi_k h_k(g(\Xc)) = \argmax_{k = 1,\ldots,K} \pi_k h_k(\theta_{(\Xc)}).
\end{align}
In the latter case, a classifier based on $\Xc$ is as good as a classifier based on the parameter $\theta_{(\Xc)}$ of the set.
This theoretical observation, as well as the empirical observation from the real data in Fig. 2, has contributed in considering a classification based on the parameter $\theta$ (or the features of a set) rather than individual observations in $\Xc$.

The proposed set classification procedure can be thought of as building a classification rule on the predicted parameter (feature) $\hat\theta$. The parameter is modeled to include the mean and PC spaces. A remaining question is the quality of the prediction of $\theta$. If $\hat\theta$ is close to $\theta$, then the empirical classifier is as good as the optimal classifier (\ref{eq:hierrisk2opt}). In the next section we show that the empirical PC space features are as good as the true PC spaces in the high dimensional asymptotic context.


\subsection{High dimensional asymptotic theory}
The quality of the empirical features $\Lc^{(r)}_i$ is now assessed in the high-dimensional, low-sample size (HDLSS) asymptotic context to support the theoretical motivation in Section~\ref{sec:Hierarchical model}. Our image dataset suits well in HDLSS context since it is in very high dimension with $p$ being tens of thousands while the number of sets $N$ and the number of observations $n$ in the set  are both small. An asymptotic investigation where $p \to \infty$ and both $(N, n)$ fixed is thus relevant to the analysis of such HDLSS data \citep{Hall2005, Ahn2007}.

Since we are interested in the empirical PC spaces, assume for $X_{ij(p)}$, the $j$th random observation in the $i$th set, $\E ( X_{ij(p)}) = \0v_p$ and
$\Cov (X_{ij(p)}) = \Sigma_{i(p)} = \sum_{l = 1}^p \lambda_{li(p)}^o e_{li(p)}^o (e_{li(p)}^o)^T $. Here, the subscript $(p)$ is used to emphasize the dependence on the dimension $p$. Let $\Uc^{(r)}_{i(p)} = \mbox{span}\{e_{1i(p)}^o,\ldots, e_{ri(p)}^o\}$ be the true PC space, which is the target of the empirical PC space $\Lc^{(r)}_i \equiv \Lc^{(r)}_{i(p)}$.

The following theorem states that when the first $r$ eigenvalues are much larger than the others, then the set classification based on the empirical PC spaces $\Lc^{(r)}_{i(p)}$ is as good as using the true PC spaces $\Uc^{(r)}_{i(p)}$ in the limit.

For two sequences $a_p$ and $b_p$ let $a_p \asymp b_p$ stand for the relation $\lim_{p\to\infty}{a_p/b_p} \in (0,\infty)$ and $\lim_{p\to\infty}{a_p/b_p} \in (0,\infty)$.

We need the following moment condition:

(C)  Each components of $X_{ij(p)}$ have finite fourth moments for all $p$, and there exists a permutation so that the permuted sequence of loadings of the scaled PCs $Z_{ij(p)} = ((\lambda_{li(p)}^o)^{-\half} (e_{li(p)}^o)^T X_{ij(p)})_{l = 1,\ldots, p}$ are $\rho$-mixing \citep{Bradley2005}.

\begin{theorem}\label{thm:hdlssconsistency}
Assume that $\lambda_{li(p)}^o  \asymp p^\alpha$, for $l = 1,\dots,r_0$ and the rest of eigenvalues are fixed, for example $\lambda_{li(p)}^o \equiv 1$ for $l = r_0 + 1, \ldots, p$. Suppose $n$ is fixed and (C)  holds.
\begin{enumerate}
  \item[(a)] If $\alpha > 1$, then $\Lc^{(r)}_{i(p)}$ is consistent with $\Uc^{(r)}_{i(p)}$ in the sense that
$\rho_s(\Lc^{(r)}_{i(p)},\Uc^{(r)}_{i(p)} ) \to 0$ in probability  as $p \to \infty$.
  \item[(b)] If $\alpha < 1$, then $\Lc^{(r)}_{i(p)}$ is strongly inconsistent with  $\Uc^{(r)}_{i(p)}$ in the limit $p \to \infty$ in the sense that
$\rho_s(\Lc^{(r)}_{i(p)},\Uc^{(r)}_{i(p)}) \to \sqrt{r} \mbox{ in probability  as }  p \to \infty.$
  \item[(c)] If $\alpha = 1$, $\rho_s(\Lc^{(r)}_{i(p)},\Uc^{(r)}_{i(p)})$ weakly converges to a distribution with support on $[0,\sqrt{r}]$ as $p \to \infty$.
\end{enumerate}
\end{theorem}
Theorem \ref{thm:hdlssconsistency} tells us that, in the limit $p \to \infty$, $\Lc^{(r)}_{i(p)}$ can be used in place of $\Uc^{(r)}_{i(p)}$ when signal from the PCs is strong (\textit{i.e.}, large $\lambda_{li(p)}^o$). Theorem~\ref{thm:hdlssconsistency} can be shown by an application of Theorem 2 of \cite{Jung2009a} (for $\alpha \neq 1$) and Theorem 3 of \cite{Jung2012} (for $\alpha = 1$) with the fact that $\sup_{U_r,V_r}\rho(U_r, V_r) = \sqrt{r}$ for $U_r,V_r \in  G(r,p)$.

Next, we show that the classification based on $\Uc^{(r)}_{i(p)}$ is optimal in a simplified setting. Suppose that in the binary classification we have;
\begin{assumption}\label{assume1}
$\Cov(X_{ij(p)}) = \Sigma_{i(p)}$ is modeled with a single factor as $\Sigma_{i(p)} = \Id_p + \uv_{i(p)} \uv_{i(p)}^T$, where
$\uv_{i(p)} \mid (Y= +1) \sim N_p(\muv_+(p), \Id_p)$ and
$\uv_{i(p)} \mid (Y= -1) \sim N_p(\muv_-(p), \Id_p)$.
\end{assumption}
The following theorem is the result of directly applying the geometric representation theory in \citet{Hall2005,Ahn2007} and \citet{Qiao2010Weighted}. Note that we have assumed normality for $\uv_{i(p)}$ to simplify the situation.

\begin{prop}
Under Assumption \ref{assume1}, the pairwise distances between the $N_+$ ($N_-$, resp.) leading eigenvectors for sets from the positive (negative, resp.) class are approximately the same. In particular, scaled by $1/p$, the squared distances satisfy
$\frac{1}{p}\|\uv_k^+-\uv_l^+\|^2\rightarrow 2$ and $\frac{1}{p}\|\uv_k^--\uv_l^-\|^2\rightarrow 2.$
\end{prop}

Now if we further assume that the squared mean difference for the $\uv$'s is $\frac{1}{p^{\alpha}}\norm{\muv_+(p) - \muv_-(p)}^2 \rightarrow \mu^2$, then we can derive the following asymptotic result which outlines the conditions under which the SVM classifier can always correctly classify the set data based on the leading eigenvectors.

\begin{theorem}
Without loss of generality, assume that $N_+\leq N_-$, and hence $1/N_+\geq 1/N_-$.
\begin{enumerate}
		\item If $\alpha>1$, then with probability converging to 1 as $p\rightarrow \infty$, a new set from either class will be correctly classified by the SVM classifier.
		\item If $\alpha<1$, then with probability converging to 1 as $p\rightarrow \infty$, a new set from either class will be misclassified by the SVM classifier.
		\item If $\alpha=1$, then when $\mu^2>1/N_+-1/N_-$, with probability converging to 1 as $p\rightarrow \infty$, a new set from either class will be correctly classified by the SVM classifier.
\end{enumerate}
\end{theorem}

\subsection{Proof of Proposition 1}\label{sec:Proof of Proposition 1}
The identifiability of the parameter $\Lc^{o} = \mbox{span}( \ev_{1}^o,\ev_{2}^o)$ is assured by the assumption $\lambda_{1}^o = \lambda_{2}^o > \lambda_{3}^o$ and the fact that the spectral decomposition is unique up to sign change.
See, for example, \citet[p. 523]{Casella2002} for the notion of identifiability.

The consistency of $\Lc^{(2)}_i = \mbox{span}(\ev_{1 i}, \ev_{2 i})$ with $\Lc^{o}$ for the high-dimensional, low-sample size context ($p \to \infty$ with fixed $n$) is stated in Theorem~\ref{thm:hdlssconsistency}. The large sample asymptotic consistency is trivially obtained again by the uniqueness of the spectral decomposition as well as the consistency of the empirical covariance matrix $\wh\Sigma_i$ with $\Sigma_i$.

\subsection{Proof of Theorem 3}\label{sec:Proof of Theorem 2}
With the notation $\zv = (\zv_{(1)}^T ; z_{(2)})^T \in \Real^{m+1}$ , the loss function $L_N$ is written as
 \begin{align*}
 L_N( \zv) &= \norm{-\frac{1}{2} P_N [D(\Zv_\dagger(\zv)) - \Delta_\dagger] P_N^T }_F^2
           = \norm{ (\Zv_\dagger(\zv))^T \Zv_\dagger - B^\dagger}_F^2 \\
           &= \norm{\Zv^T \Zv - B}_F^2 + 2 \norm{\Zv^T\zv_{(1)}  - \bv_{12}}_F^2 + a(\zv)^2,
 \end{align*}
 where $a(\zv) = a(\zv_{(1)},z_{(2)}) = \zv_{(1)}^T\zv_{(1)} + z_{(2)}^2 - b_2$.
 The first order condition leads that
 \begin{align}\label{eq:firstorder}
 \frac{1}{4}\frac{d L_N(\zv)}{d\zv} =
 \frac{1}{4}\left( \begin{array}{c}
  \frac{dL_N(\zv)}{d\zv_{(1)}}  \\
  \frac{\partial L_N(\zv)}{\partial z_{(2)}}
 \end{array} \right)
 =
  \left[ \begin{array}{c}
      \Zv \Zv^T \zv_{(1)} - \Zv \bv_{12} + a(\zv) \zv_{(1)} \\
     2(\zv_{(1)}^T\zv_{(1)} + z_{(2)}^2 - b_2) z_{(2)}
   \end{array} \right]
  = \0v.
 \end{align}
The Hessian matrix is
\begin{align*}
H(\zv) & = \left[ \begin{array}{cc}
      \Zv \Zv^T + \zv_{(1)}\zv_{(1)}^T +a(\zv) \Id_m  & 2 z_{(2)} \zv_{(1)} \\
     2 z_{(2)} \zv_{(1)}^T &   2 a(\zv) + 4z_{(2)}^2
   \end{array} \right] \\
  & = \left[ \begin{array}{cc}
        \Lambda_+ + a(\zv)\Id_m & \0v \\
        \0v^T              & 2a(\zv)
     \end{array} \right]
     +
     \left( \begin{array}{c}
              \zv_{(1)} \\
              2z_{(2)}
          \end{array} \right)
          \left( \begin{array}{c}
                        \zv_{(1)} \\
                        2z_{(2)}
                    \end{array} \right)^T,
\end{align*}
since $\Zv \Zv^T = \Lambda_+$. Now the $\hat{\zv}$ (defined in Theorem 2) satisfies (\ref{eq:firstorder}), and $H(\hat{\zv})$ is positive definite if $\hat{z}_{(2)}^2 = b_2 - \zv_\dagger^T\zv_\dagger > 0$, because $a(\hat{\zv}) = 0$. If $\hat{z}_{(2)}^2 = 0$, then $L_N(\zv)$ is invariant to $z_{(2)}$, and $\hat{\zv}$ is in fact the global minimum.
This shows that $\hat{\zv}$ are the local minima of $L_N$.

For the second argument, if it is nonnegative definite, $B^\dagger$ can be thought of as a covariance matrix of an $(N+1)$-variate normal distribution. We have
$b_2 - \zv_\dagger^T\zv_\dagger = b_{22} - \bv_{21} B^{-1} \bv_{12} \ge 0$, since the latter is the conditional variance of the last coordinate given the first $N$ coordinates, which is nonnegative.

\newpage
\section{On the scale factor $c$ of the Euclidean Sine metric (Section 3.1.2)}\label{sec:scale factor}
\begin{figure*}[htb!]
  \begin{center}
  \ifpdf
        \includegraphics[width=0.9\textwidth,trim = 20mm 60mm 20mm 60mm]{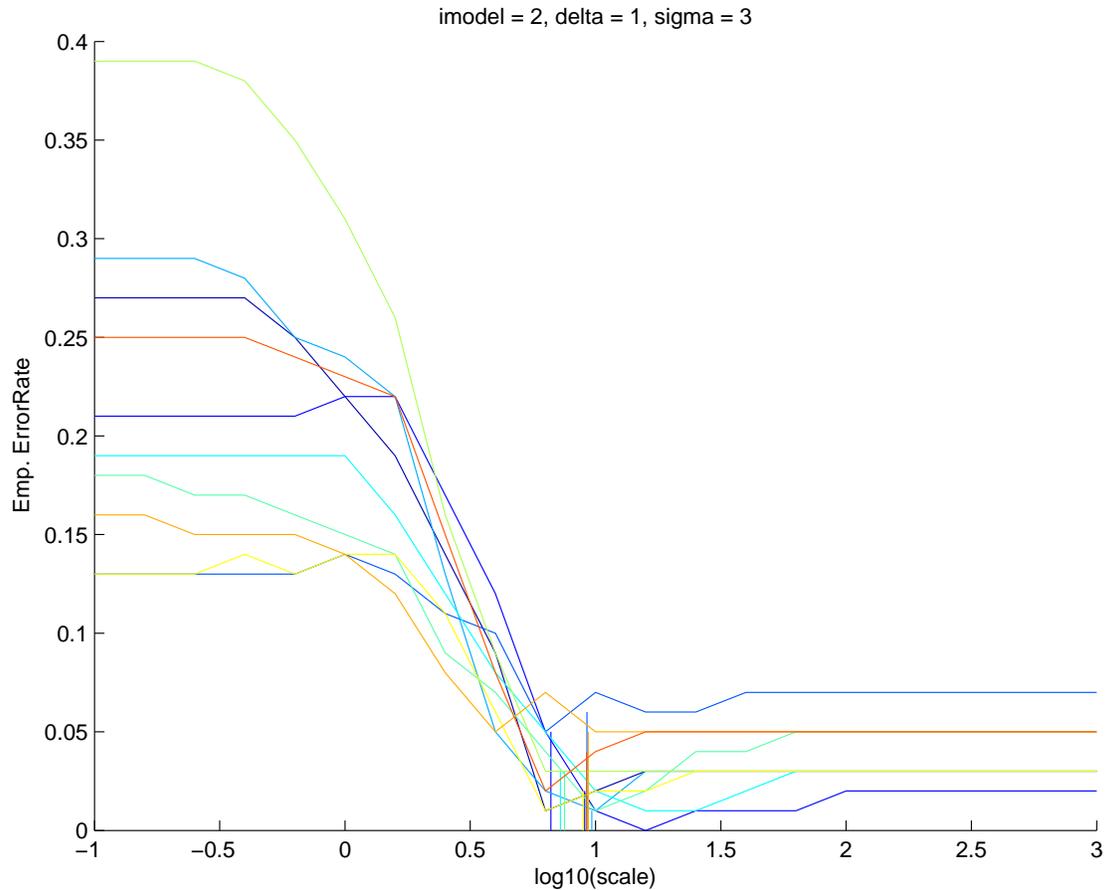}
  \else
  \fi
  \end{center}
  \caption{The choice of $c$ as the average  of the empirical total variance in $\Lc_i^{(r)}$ (our choice in Section 3.1.2) is shown to be sensible compared to other possible values of $c$. What is plotted here is the empirical error rates of PCF-LDA classifier for varied values of scale parameter $c \in (0.1, 1000)$, shown as piecewise linear curves. The computed $c$ is shown as vertical line segment. Different colors represent different realizations. Model (2) with $(p,N) = (200, 20)$ is used to generate set classification data.  \label{fig:studyonscale}}
  \end{figure*}

See Fig. \ref{fig:studyonscale}.

\newpage
\section{Comparison of different statistics (Section~4)}\label{sec:Comparison of different statistics}
We report the results of our experiments in Table~\ref{tab:1choice_of_statistic} leading that the use of $T(r)$ exhibited better performance than using other choices of statistics. For the comparison, we have considered the following statistics;
 \begin{enumerate}
   \item a modified sum of squared $t$-statistics $T_1^\tau(r)  = \etav_r' (D_r + \tau \Id_m)^{-1}\etav_r$, for $\tau \in [0, \mbox{trace}(S_r)]$ \citep{Srivastava2008},
   \item a version of  Hotelling's $T^2$ statistic $T_2^\tau(r) = \etav_r' (S_r + \tau \Id_m)^{-1}\etav_r$,
   \item Dempster's statistic $T_3(r) = \etav_r' \etav_r / \mbox{tr}(S_r)$ \citep{Dempster1960,Srivastava2006}, and
   \item the ratio of between-group and within-group distances $R_1(r) = B_1(r) / W_1(r)$, where
$$ B_1(r) = \sum_{k=1}^2 \sum_{i,j \in G_k} \rho_s(\Lc_i^{(r)},\Lc_j^{(r)}), \
 W_1(r) = \sum_{i \in G_1} \sum_{j \in G_2} \rho_s(\Lc_i^{(r)},\Lc_j^{(r)}).$$
 \end{enumerate}
 
 Table~\ref{tab:1choice_of_statistic} summarizes the performance of using these statistics in terms of the empirical error rates when $r$ is selected by the corresponding statistic, and the classification method is chosen as either LDA, QDA or the 3-Nearest Neighbors. The empirical error rates when using $\hat{r}(T)$ are significantly smaller than those from using $T_2$, $T_3$ and $R_1$, and are comparable to those from using $T_1$, for various dimension--sample size situations.

 \begin{table}
 \caption{The empirical error rates (mean and standard error based on 100 simulations) show that the performance of using $T$ is superior to using others. The last row, labeled as `Base,' is the minimum empirical errors given by exhaustive search over all $1\le r \le R$.  Model (2) in Section~5.2 is used with hyper-parameters $\delta = 1$, $\sigma = 3$, $\rho = 0$ and $\tau_s = t_s\mbox{tr}(S_r)$,  $\{t_s\}_{s=1}^3 = \{0.01, 0.1, 0.5\}$.}\label{tab:1choice_of_statistic}
\begin{center}
 \begin{tabular}{ccccccc}
  &  \multicolumn{3}{c}{$(p,N) = (20,10)$}
   &  \multicolumn{3}{c}{$(p,N) = (100,10)$} \\
 (\%)  &  LDA & QDA & 3-NN  &  LDA & QDA & 3-NN \\
   \hline
 $ T $           &  6.92(0.54)&	6.02(0.44)&	 9.32(0.63)& 6.74(0.49)&	 6.65(0.45)&	13.05(0.85)  \\
 $ T_1^{\tau_1}$ &  5.76(0.48)&	5.09(0.40)&	 8.04(0.54)& 5.81(0.43)&	 5.82(0.41)&	11.68(0.79)   \\
 $ T_1^{\tau_2}$ &  4.52(0.37)&	4.34(0.38)&	 6.88(0.46)& 5.11(0.37)&	 5.20(0.37)&	 8.68(0.48)   \\
 $ T_1^{\tau_3}$ &  6.28(0.44)&	6.16(0.45)&	 9.62(0.55)& 6.85(0.53)&	 6.94(0.53)&	10.41(0.65) \\
 $ T_2^{\tau_1}$ & 12.23(0.50)&	9.98(0.40)&	14.93(0.64)&15.56(0.91)&	 14.35(0.79)&	23.08(1.27)\\
 $ T_3 $         & 12.23(0.50)&	9.98(0.40)&	14.93(0.64)&15.95(0.90)&	 14.70(0.79)&	23.47(1.27)\\
 $ R_1$          & 12.23(0.50)&	9.98(0.40)&	14.93(0.64)&16.10(0.89)&	 14.85(0.78)&	23.64(1.25)\\
  Base           &  2.25(0.18)&	2.06(0.18)&	 3.87(0.24)& 2.45(0.16)&	 2.47(0.15)&   5.56(0.29)\\
   \hline
\end{tabular}
\bigskip \\
 \begin{tabular}{cccc}
   &  \multicolumn{3}{c}{$(p,N) = (200,20)$} \\
 (\%)  &  LDA & QDA & 3-NN   \\
   \hline
 $ T $           &3.20(0.23)&	3.22(0.24)&	10.45(0.67)\\
 $ T_1^{\tau_1}$ &3.02(0.21)&	3.02(0.22)&	 9.80(0.56) \\
 $ T_1^{\tau_2}$ &2.53(1.94)&	2.52(0.19)&	 8.15(0.35) \\
 $ T_1^{\tau_3}$ &4.30(0.48)&	4.37(0.49)&	10.42(0.63)\\
 $ T_2^{\tau_1}$ & 8.04(0.64)&	7.70(0.59)&	17.85(1.20)\\
 $ T_3 $         & 8.41(0.65)&	8.05(0.59)&	18.33(1.24)\\
 $ R_1$          & 8.44(0.65)&	8.07(0.59)&	18.33(1.24)\\
  Base           &  1.67(0.14)&	1.66(0.14)&	 6.85(0.33)\\
   \hline
\end{tabular}
\end{center}
\end{table}

\section{Validation for the permutation test (Section~4)}\label{sec:Validatio}

\begin{figure*}[htb!]
\begin{center}
\ifpdf
      \includegraphics[width=0.7\textwidth,trim = 20mm 80mm 20mm 80mm]{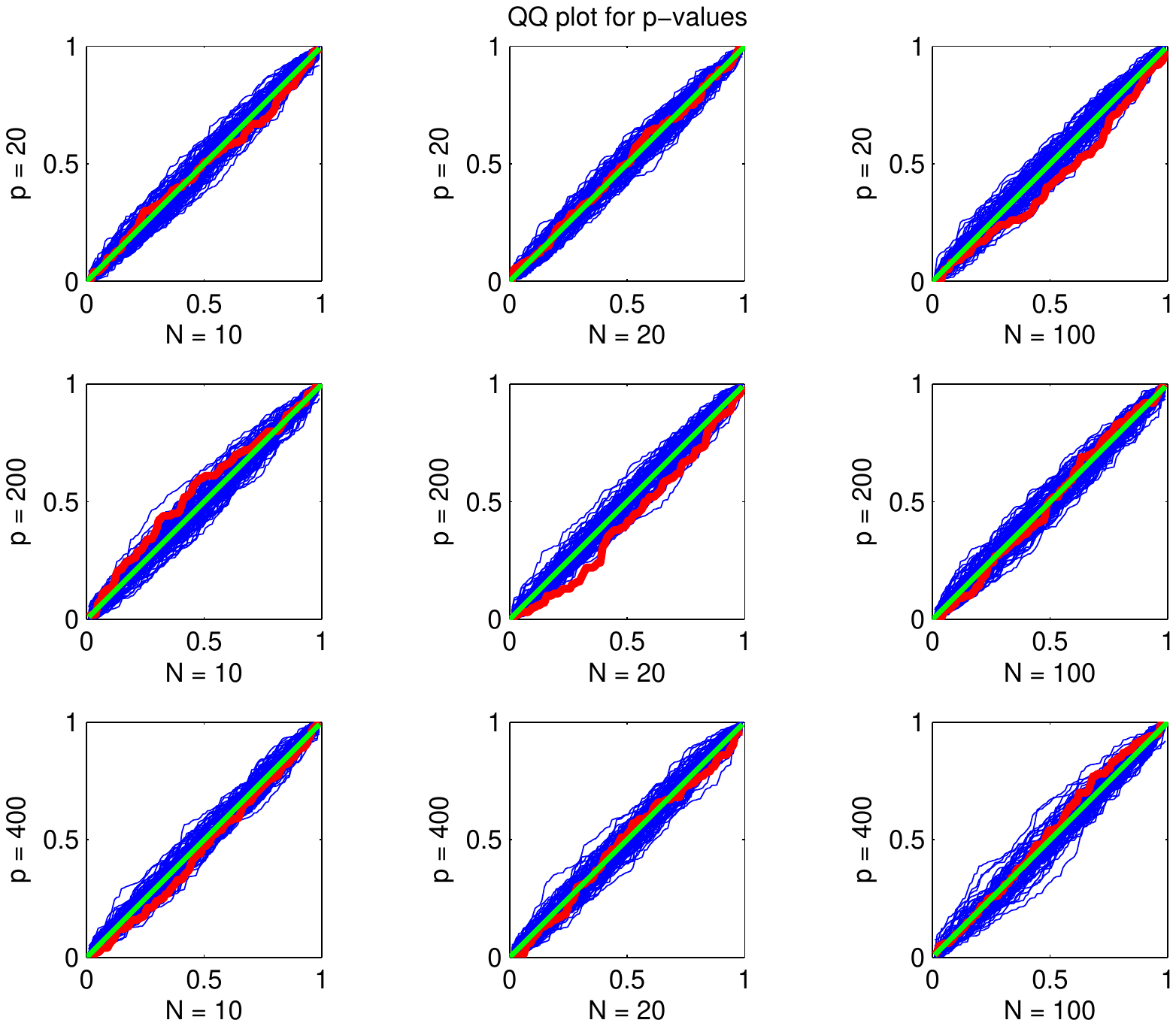}
\else
\fi
\end{center}
\caption{Quantile-quantile (QQ) envelop plot of the empirical p-values under the null hypothesis of the permutation test proposed in Section~4.2. The quantile-quantile (QQ) plots (shown as red curves) of the empirical p-values, in nine difference combinations of dimensions and sample sizes $(p = 10,100, 400, N = 10, 20, 100)$ are shown. We use the QQ envelope plot to understand the natural variation of the empirical quantiles \citep{Lee2007}. The green line in each panel of Figure~\ref{fig:permtest_qqplot} shows the theoretical quantiles from the uniform$(0,1)$ distribution. Hundred QQ plots from random samples of the same size are overlaid as blue curves. The red QQ curves are inside of the blue curve bundles, from which we conclude that the empirical p-values follow the uniform$(0,1)$ distribution.   \label{fig:permtest_qqplot}}
\end{figure*}

\begin{figure*}[htb!]
 \begin{center}
 \ifpdf
       \includegraphics[width=0.7\textwidth,trim = 20mm 80mm 20mm 60mm]{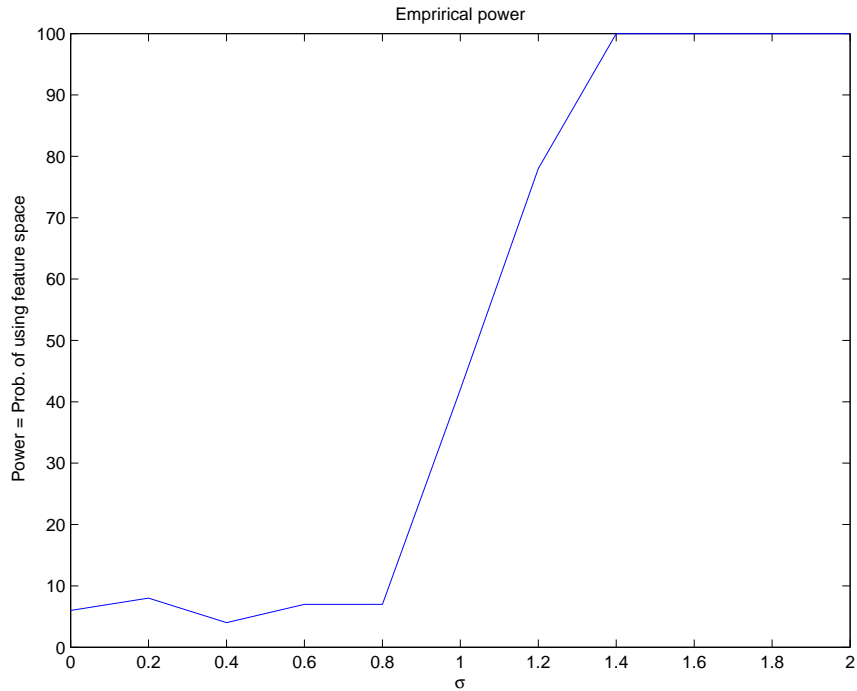}
 \else
 \fi
 \end{center}
 \caption{Empirical power function of the permutation test (Section~4.2) on different signal strength $0 \le \sigma \le 2$. Model (2) with $(p,N) = (10,20)$ is used.   \label{fig:permtest_power}}
 \end{figure*}

\begin{figure*}[htb!]
  \begin{center}
  \ifpdf
        \includegraphics[width=0.9\textwidth,trim = 20mm 60mm 20mm 60mm]{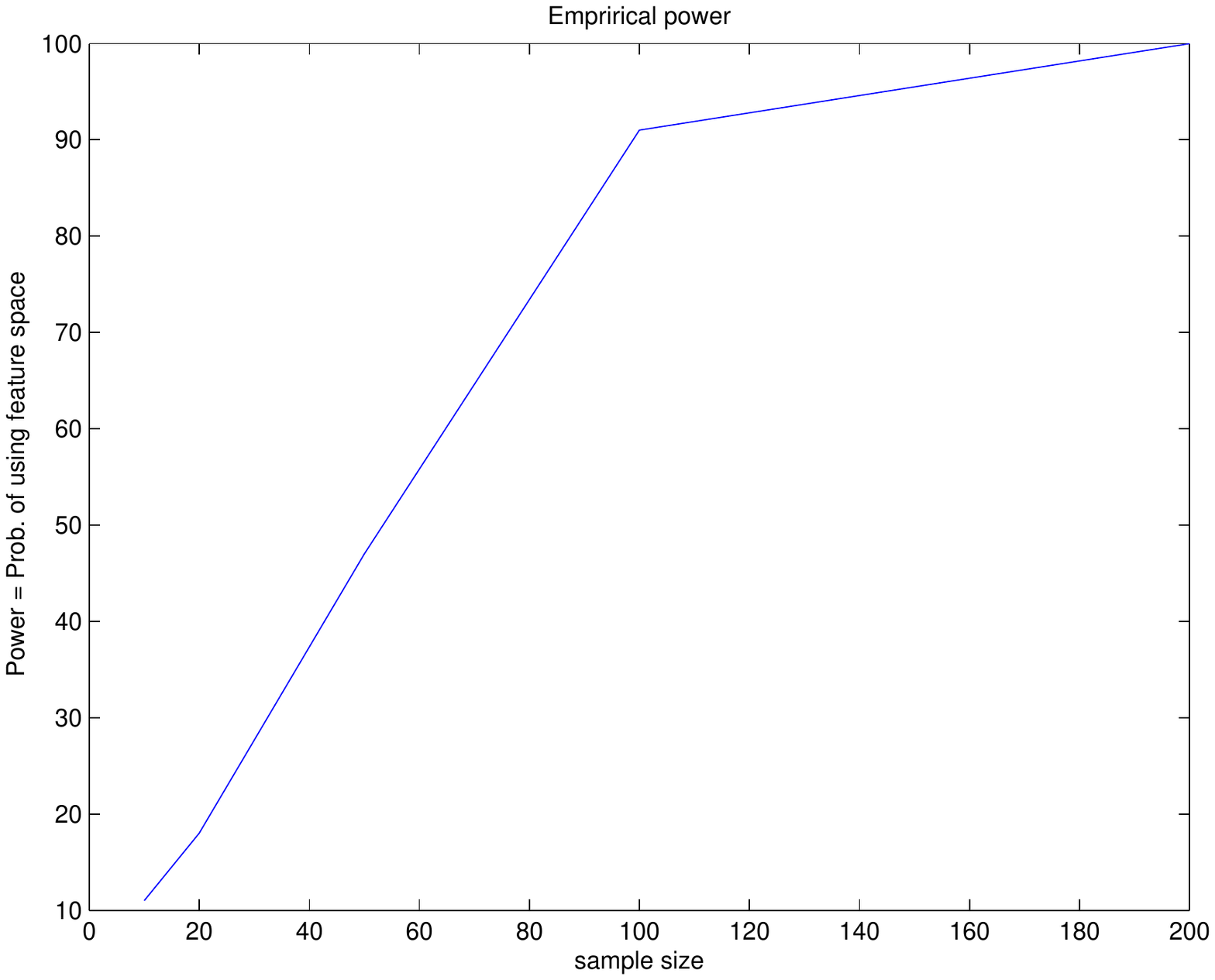}
  \else
  \fi
  \end{center}
  \caption{Empirical power function of the permutation test (Section~4.2) on different sample size $N = 10, 20, 50, 100, 200$ (the number of sets). Model (2) with $p = 10$, $\sigma = 1$ is used.   \label{fig:permtest_power_s}}
  \end{figure*}

We first demonstrate that the proposed permutation test leads to uniform null distribution for p-values. we report in Fig.~\ref{fig:permtest_qqplot} using an envelop QQ plot \citep{Lee2007}.
The empirical power of the proposed permutation test is also examined. The power increases with the signal strenth $\sigma$, and also with the sample size $N$, as shown in Figures~\ref{fig:permtest_power} and \ref{fig:permtest_power_s}. We used $p=10$, $\sigma = 0, 1/2, 1, 3/2, 2$ with fixed $N = 20$ with the model (2) (\emph{cf}. Section 5) for Figure~\ref{fig:permtest_power}, and $\sigma = 1$ with $N = 10, 20, 50, 100, 200$ for Figure~\ref{fig:permtest_power_s}. Note also that the power also increases with the number of instances $n_i$ grows.

 When the true underlying model has no covariance difference between groups, the proposed permutation test greatly enhances the performance. Table \ref{tab:4Model1} compares the classification results before (using $\hat{r} > 0$) and after updating $\hat{r}$ by the decision of the test. The empirical error rates are computed using LDA, while using other classifiers exhibits similar behavior. The error rates are notably decreased after applying the permutation test. The target $r$ is zero because we used a hierarchical model with no covariance difference. In this experiment, the $\hat{r}$ is updated to zero in 92\% out of 100 simulations (when $T$ is used). Performances of using different statistics are also summarized in Table \ref{tab:4Model1}. While the benefit of using the test is clear, different choices of test statistic do not significantly affect the results.

\begin{table}
 \caption{The empirical error rates (mean and standard error based on 100 simulations) are significantly reduced when the permutation test is conducted. The true underlying model has no covariance difference (Model (1) in Section~5.2. \label{tab:4Model1}}
 \begin{center}
 \begin{tabular}{ccccccc}
    &  \multicolumn{2}{c}{$(p,N) = (20,10)$}
   &  \multicolumn{2}{c}{$(p,N) = (100,10)$}
   &  \multicolumn{2}{c}{$(p,N) = (200,20)$} \\
 $(\%)$  &  before & after  & before & after & before & after \\
   \hline
 $ T $          &18.68(1.12)&       7.38(0.49)& 24.48(1.20)& 14.20(0.74)& 21.75(1.00)& 13.83(0.62)\\
 $ T_1^{\tau_1}$&21.13(1.09)&       7.38(0.50)& 26.14(1.17)& 14.20(0.74)& 25.79(0.95)& 13.83(0.62)\\
 $ T_1^{\tau_2}$&26.02(0.84)&       7.93(0.60)& 34.89(0.84)& 15.04(0.90)& 35.03(0.61)& 14.34(0.71)\\
 $ T_1^{\tau_3}$&29.16(0.66)&       8.63(0.74)& 40.45(0.59)& 14.84(0.88)& 36.80(0.59)& 14.08(0.67)\\
 $ T_2^{\tau_1}$&18.14(1.25)&       8.35(0.73)& 22.70(1.22)& 13.19(0.53)& 17.58(0.86)& 12.95(0.45)\\
 $ T_3 $        &18.19(1.25)&       8.07(0.69)& 22.85(1.23)& 13.20(0.53)& 17.47(0.86)& 12.95(0.45)\\
 $ R_1$         &18.19(1.25)&       8.07(0.69)& 22.90(1.22)& 13.20(0.53)& 17.47(0.86)& 12.95(0.45)\\
 Exhaustive     &  \multicolumn{2}{c}{  4.47(0.27)}&  \multicolumn{2}{c}{ 11.45(0.49)}&  \multicolumn{2}{c}{ 10.94(0.36)}\\
 ($r = 0$)     &  \multicolumn{2}{c}{  7.58(0.47)}&  \multicolumn{2}{c}{  13.74(0.42)}&  \multicolumn{2}{c}{  12.79(0.42)}\\
 \hline
 \end{tabular}
 \end{center}
 \bigskip
\end{table}
 
\newpage

\section{Definitions of modified linear discriminant analyses (Section 5)}
\label{sec:LDAdefs}
 The LDA and QDA used in the simulation studies and real data analysis are modified from the usual definition, for use in the high-dimensional, low-sample size situations. The LDA is notorious for its poor performance when $p \ge n$ \citep{Bickel2004,Ahn2010}. Both classification rules are dependent on the (pooled) sample covariance matrices. We simply replace the sample covariance matrix $\widehat\Sigma$ with a shrinkage estimator $\widehat\Sigma + \gamma \Id$ for a small positive constant $\gamma$. We have used $\gamma = 0.01$ for all analyses in the paper. The idea of modified discriminant analysis dates back to \cite{Friedman1989}.

An anonymous referee has suggested to compare the performances of the proposed method with those of the modified linear discriminant analyses appeared in \cite{Srivastava2007} and \cite{Yata2012}. Here precise definitions of those methods are given.

Consider binary classification with $Y = \pm 1$. With the pooled sample covariance matrix $\Sv$ and the sample group means $\hat\mu_+$ and $\hat\mu_-$, Fisher's linear discriminant function is, for a new observation $\xv$,
\begin{equation}\label{fisher'slda}
\delta(\xv) = \sgn (\xv - \frac{\hat\mu_+ + \hat\mu_-}{2} )^T\Sv^{-1}(\hat\mu_+ - \hat\mu_-).
\end{equation}
As discussed in Section 5.1, use of a modified linear discriminant function is a common practice, by replacing $\Sv$ by  $\Sv + \gamma \Iv_p$, for an appropriate value of $\gamma$. We have used
$$ \delta_\gamma(\xv) = \sgn (\xv - \frac{\hat\mu_+ + \hat\mu_-}{2} )^T(\Sv + \gamma \Iv_p)^{-1}(\hat\mu_+ - \hat\mu_-),$$
with $\gamma = 0.01$.

\cite{Srivastava2007} have compared several linear discriminant methods and concluded that minimum distance empirical Bayes rule (MDEB) performs best among all methods considered under the high-dimensional, low-sample-size situation. The MDEB is obtained by choosing $\gamma =  \tr(\Sv)/ \min(n,p)$, so that the classification function is
$$ \delta_{MDEB}(\xv) = \sgn (\xv - \frac{\hat\mu_+ + \hat\mu_-}{2} )^T(\Sv + \frac{\tr(\Sv)}{\min(n,p)} \Iv_p)^{-1}(\hat\mu_+ - \hat\mu_-).$$

\cite{Yata2012} have considered a hard-thresholded covariance estimate $\Sv_\omega$ to replace $\Sv$ in (\ref{fisher'slda}). Denote the eigen-decomposition of the sample covariance matrix as $\Sv = \sum_{j=1}^{p}\lambda_j \ev_j \ev_j^T$, where $\lambda_1 > \lambda_2 > \cdots > \lambda_{n-1}>0$ and $\lambda_{j} = 0$ for  $j = n, \ldots,p$. For $$\omega = \min\{ \frac{\tr (\Sv) }{p^{1/2}n^{1/4}} ,\frac{\tr(\Sv)}{\min(n,p)}\},$$ the hard-thresholded covariance estimate of \cite{Yata2012} is defined as
$$ \Sv_\omega = \sum_{j=1}^{n-1} \max\{\lambda_j, \omega \left( \frac{\lambda_j}{\lambda_j - \frac{\sum_{i = j+1}^{n-1} \lambda_j}{n-j}} \right)\} \ev_j\ev_j^T + \sum_{j=n}^{p} \omega \ev_j\ev_j^T.$$
The corresponding classification function is
$$ \delta_{YA}(\xv) = \sgn (\xv - \frac{\hat\mu_+ + \hat\mu_-}{2} )^T\Sv_\omega^{-1}(\hat\mu_+ - \hat\mu_-).$$

The majority voting classifiers, denoted as 'Classifier'-MV, are constructed as a majority vote of the $\delta(\xv_i)$, $i = 1,\ldots,n$; for example,
$$ \delta_{MDEB-MV}(\{\xv_i\}) = \sgn \left( \sum_{i=1}^n \sgn (\xv - \frac{\hat\mu_+ + \hat\mu_-}{2} )^T\Sv_\omega^{-1}(\hat\mu_+ - \hat\mu_-) \right).$$  
The weighted voting classifier is defined by 
$$ \delta_{MDEB-WV}(\{\xv_i\}) = \sgn \left( \sum_{i=1}^n (\xv - \frac{\hat\mu_+ + \hat\mu_-}{2} )^T\Sv_\omega^{-1}(\hat\mu_+ - \hat\mu_-) \right).$$

%

\section{Additional tables for Section~5 (Numerical studies)}\label{sec:numstudiesTables}
See Tables \ref{tab:simulationResultsModelFULL} and \ref{tab:simulationResultsModelFULL2}.
\begin{table}
\caption{Empirical misclassification rates ($\rho = 0$).}\label{tab:simulationResultsModelFULL}
\begin{footnotesize}
\begin{tabular}{ccccccc}
\multicolumn{7}{c}{Model (1) with $\rho = 0$}\\
($p, N)$  & (20,10) & (200, 10) & (400,10)& (20,20) & (200,20) & (400,20)\\
\hline
    PCF-LDA & 25.41(  5.43) & 33.13(  3.49) & 36.57(  3.88) & 20.15(  5.01) & 28.34(  3.82) & 33.18(  3.86) \\
    PCF-QDA & 29.07(  6.00) & 34.11(  3.99) & 37.06(  3.80) & 32.62(  6.05) & 31.84(  3.94) & 34.84(  3.85) \\
    PCF-SVM & 22.03(  5.45) & 33.10(  3.51) & 36.61(  3.88) & 20.74(  5.08) & 28.30(  3.81) & 33.22(  3.89) \\
    PCF-DWD & 18.75(  4.28) & 32.74(  3.73) & 36.60(  4.01) & 15.29(  3.68) & 26.74(  3.34) & 32.61(  3.97) \\
    PCF-MDEB & 18.26(  4.42) & 32.59(  3.82) & 36.62(  4.02) & 14.81(  3.65) & 26.65(  3.41) & 32.59(  3.93) \\
    PCF-YA & 21.32(  7.64) & 46.87(  5.52) & 47.65(  3.84) & 14.23(  3.61) & 41.38(  5.97) & 45.88(  4.34) \\
    LDA-MV & 20.21(  4.64) & 38.46(  5.01) & 44.53(  3.89) & 16.63(  4.14) & 31.43(  4.64) & 38.69(  4.10) \\
    LDA-WV & 18.18(  4.60) & 37.57(  4.91) & 44.23(  3.80) & 14.29(  3.89) & 29.44(  4.92) & 37.61(  4.47) \\
    QDA-MV & 27.93(  5.84) & 49.63(  1.32) & 49.26(  2.00) & 21.40(  5.30) & 47.39(  2.96) & 49.48(  1.53) \\
    QDA-WV & 26.31(  5.77) & 49.43(  1.75) & 49.41(  1.56) & 19.38(  5.35) & 47.12(  2.81) & 49.57(  1.46) \\
    SVM-MV & 20.72(  4.56) & 38.27(  4.36) & 40.70(  4.13) & 16.82(  4.29) & 32.39(  4.16) & 38.27(  4.70) \\
    DWD-MV & 19.55(  4.14) & 34.06(  3.50) & 37.58(  3.70) & 16.12(  3.84) & 28.30(  3.73) & 34.09(  3.84) \\
    MDEB-MV & 21.11(  5.95) & 35.18(  3.86) & 38.86(  4.08) & 17.10(  4.29) & 28.99(  4.11) & 35.20(  4.23) \\
    MDE-WV & 19.57(  6.08) & 33.69(  4.37) & 37.93(  4.29) & 14.71(  4.07) & 27.63(  4.20) & 33.42(  4.01) \\
    \hline
\multicolumn{7}{c}{Model (2) with $\rho = 0$}\\
 \hline
    PCF-LDA &4.06(  3.58) & 2.35(  3.84) & 5.38(  4.00) & 0.85(  0.85) & 0.79(  0.71) & 2.22(  1.25) \\
    PCF-QDA &7.32(  5.79) & 2.30(  3.77) & 5.34(  4.00) & 3.82(  3.77) & 0.72(  0.69) & 2.16(  1.24) \\
    PCF-SVM &5.20(  4.39) & 2.35(  3.84) & 5.38(  4.00) & 1.57(  2.00) & 0.79(  0.71) & 2.22(  1.25) \\
    PCF-DWD &6.56(  4.76) & 2.35(  3.93) & 5.38(  4.00) & 3.01(  2.85) & 0.77(  0.68) & 2.22(  1.24) \\
    PCF-MDEB & 6.14(  4.60) & 2.36(  3.87) & 5.38(  4.01) & 2.20(  2.36) & 0.77(  0.69) & 2.22(  1.26) \\
    PCF-YA & 9.12(  7.52) & 22.47( 10.64) & 30.61(  8.84) & 1.50(  1.59) & 19.39( 10.66) & 26.43(  9.89) \\
    LDA-MV & 39.78(  4.26) & 47.69(  3.53) & 46.24(  3.40) & 36.41(  4.22) & 45.91(  3.49) & 47.60(  3.36) \\
    LDA-WV & 38.18(  4.72) & 47.48(  3.49) & 45.92(  3.44) & 34.79(  4.42) & 45.27(  3.94) & 47.80(  3.66) \\
    QDA-MV & 4.71(  2.14) & 25.73(  3.78) & 24.09(  3.17) & 2.25(  1.24) & 35.07(  4.25) & 29.86(  3.61) \\
    QDA-WV & 2.04(  1.49) & 21.96(  3.84) & 20.10(  2.99) & 0.63(  0.64) & 33.36(  4.93) & 27.40(  3.71) \\
    SVM-MV & 39.95(  4.80) & 46.08(  3.45) & 44.99(  3.54) & 36.38(  4.03) & 46.48(  3.86) & 46.55(  3.37) \\
    DWD-MV & 36.74(  4.61) & 38.75(  4.19) & 40.39(  4.07) & 34.77(  4.08) & 36.28(  3.80) & 38.93(  3.41) \\
    MDEB-MV & 36.16(  6.28) & 42.81(  3.64) & 43.30(  4.03) & 32.23(  5.47) & 41.09(  3.88) & 43.46(  3.89) \\
    MDE-WV & 34.34(  7.26) & 41.49(  3.92) & 42.77(  3.99) & 30.29(  6.07) & 39.91(  4.23) & 43.31(  3.31) \\
    \hline
\multicolumn{7}{c}{Model (3) with $\rho = 0$}\\
 \hline
    PCF-LDA &5.80(  4.07) & 17.90(  6.87) & 27.88(  7.40) & 2.84(  1.89) & 12.05(  5.52) & 21.03(  6.94) \\
    PCF-QDA &6.25(  4.53) & 18.37(  7.12) & 28.19(  7.27) & 4.14(  3.15) & 13.99(  6.58) & 22.71(  7.60) \\
    PCF-SVM &5.75(  3.83) & 17.89(  6.86) & 27.40(  5.78) & 3.17(  2.18) & 11.97(  5.43) & 21.03(  6.94) \\
    PCF-DWD &6.10(  3.79) & 18.10(  6.58) & 28.14(  4.86) & 3.59(  2.55) & 11.97(  5.15) & 20.88(  6.74) \\
    PCF-MDEB& 7.34(  5.69) & 20.00(  6.90) & 30.10(  6.38) & 3.49(  2.46) & 12.50(  4.82) & 23.35(  7.18) \\
    PCF-YA & 11.44(  8.05) & 41.13(  8.31) & 44.94(  7.57) & 3.61(  2.76) & 39.85(  6.84) & 44.71(  6.09) \\
    LDA-MV & 36.45(  6.03) & 36.31(  4.42) & 36.27(  4.29) & 35.04(  5.54) & 35.61(  4.58) & 34.94(  3.58) \\
    LDA-WV & 36.10(  6.33) & 35.92(  4.36) & 36.25(  4.27) & 34.38(  5.85) & 35.52(  4.53) & 34.67(  3.48) \\
    QDA-MV & 27.64(  4.25) & 32.46(  3.33) & 34.20(  3.96) & 28.66(  6.94) & 33.78(  3.65) & 34.89(  3.34) \\
    QDA-WV & 25.59(  4.27) & 31.28(  3.30) & 33.24(  3.88) & 26.29(  6.85) & 32.28(  3.39) & 34.01(  3.59) \\
    SVM-MV & 36.74(  6.14) & 33.48(  4.45) & 34.74(  4.41) & 33.13(  6.32) & 32.75(  4.93) & 32.57(  3.66) \\
    DWD-MV & 29.35(  6.50) & 28.72(  4.04) & 31.80(  4.26) & 28.99(  6.07) & 24.65(  3.80) & 27.87(  3.57) \\
    MDEB-MV  &35.61( 11.43) & 31.01(  6.47) & 33.96(  5.81) & 37.45( 12.18) & 26.94(  6.37) & 30.39(  6.59) \\
    MDE-WV & 34.81( 12.20) & 30.52(  6.48) & 33.74(  5.86) & 36.89( 12.75) & 26.30(  6.50) & 30.14(  6.78) \\
    \hline
\multicolumn{7}{c}{Model (4) with $\rho = 0$}\\
 \hline
    PCF-LDA &13.78(  5.66) & 33.40(  4.06) & 38.07(  4.07) & 1.53(  1.56) & 30.31(  4.03) & 32.49(  3.58) \\
    PCF-QDA &21.14(  5.68) & 33.50(  3.81) & 38.38(  4.15) & 17.10(  4.75) & 33.20(  4.25) & 34.75(  3.80) \\
    PCF-SVM &15.53(  5.41) & 33.42(  4.03) & 38.09(  4.05) & 4.95(  4.14) & 30.27(  4.02) & 32.50(  3.55) \\
    PCF-DWD &15.73(  4.81) & 32.75(  4.28) & 37.82(  4.15) & 10.21(  3.54) & 29.02(  3.78) & 31.83(  3.56) \\
    PCF-MDEB &15.36(  4.55) & 32.65(  4.19) & 37.80(  4.05) & 8.80(  3.77) & 28.91(  3.77) & 31.65(  3.53) \\
    PCF-YA & 16.57(  8.63) & 45.65(  6.03) & 48.23(  4.64) & 0.54(  0.97) & 42.89(  5.52) & 46.91(  4.51) \\
    LDA-MV & 38.90(  4.02) & 41.60(  3.85) & 45.82(  3.87) & 35.52(  4.39) & 38.02(  4.10) & 39.27(  4.41) \\
    LDA-WV & 37.25(  4.36) & 41.12(  4.21) & 45.45(  4.11) & 33.84(  4.70) & 37.50(  4.11) & 38.24(  4.94) \\
    QDA-MV & 4.34(  1.99) & 49.68(  1.46) & 49.23(  1.99) & 2.72(  1.54) & 47.28(  2.96) & 49.61(  1.84) \\
    QDA-WV & 1.02(  1.25) & 49.57(  1.57) & 49.53(  1.66) & 0.23(  0.35) & 47.02(  3.30) & 49.63(  1.40) \\
    SVM-MV & 38.28(  4.20) & 41.38(  3.98) & 41.50(  3.89) & 35.59(  4.15) & 39.51(  4.36) & 38.72(  4.39) \\
    DWD-MV & 36.71(  3.97) & 35.00(  4.04) & 38.76(  3.60) & 34.04(  4.20) & 31.63(  3.43) & 33.51(  3.50) \\
    MDEB-MV & 35.58(  5.36) & 37.60(  3.87) & 40.22(  3.83) & 32.51(  5.49) & 34.66(  3.80) & 34.83(  3.91) \\
    MDE-WV & 33.83(  6.30) & 36.39(  3.89) & 39.48(  3.76) & 30.23(  6.16) & 33.20(  4.01) & 33.55(  3.92) \\
     \hline
\end{tabular}
\end{footnotesize}
\end{table}

\begin{table}
\caption{Empirical misclassification rates ($\rho = 0.5$).}\label{tab:simulationResultsModelFULL2}
\begin{footnotesize}
\begin{tabular}{ccccccc}
\multicolumn{7}{c}{Model (1) with $\rho = 0.5$}\\
($p, N)$  & (20,10) & (200, 10) & (400,10)& (20,20) & (200,20) & (400,20)\\
\hline
    PCF-LDA & 23.69(  5.49) & 33.76(  4.58) & 38.77(  4.38) & 18.60(  4.52) & 28.53(  4.68) & 33.93(  4.75) \\
    PCF-QDA & 27.15(  5.66) & 36.42(  4.25) & 40.14(  4.31) & 30.98(  6.45) & 33.52(  4.44) & 37.46(  4.30) \\
    PCF-SVM & 20.57(  4.96) & 33.73(  4.58) & 38.75(  4.38) & 19.03(  4.99) & 28.34(  4.68) & 34.01(  4.81) \\
    PCF-DWD & 17.44(  3.96) & 34.31(  4.72) & 39.58(  4.42) & 14.27(  3.46) & 27.78(  4.98) & 34.28(  4.76) \\
    PCF-MDEB & 17.09(  3.99) & 34.34(  4.82) & 40.04(  4.58) & 13.94(  3.31) & 27.94(  4.96) & 34.41(  4.82) \\
    PCF-YA & 19.35(  6.27) & 45.50(  5.28) & 48.62(  4.02) & 16.16(  4.89) & 42.35(  5.91) & 46.80(  4.08) \\
    LDA-MV & 18.08(  3.99) & 36.82(  5.32) & 43.51(  3.91) & 14.53(  3.86) & 29.71(  5.01) & 37.13(  4.41) \\
    LDA-WV & 16.56(  4.07) & 36.17(  5.08) & 43.17(  4.06) & 12.99(  3.70) & 28.43(  5.02) & 36.13(  4.51) \\
    QDA-MV & 24.11(  4.96) & 49.19(  1.94) & 48.87(  2.49) & 18.54(  4.97) & 45.96(  3.24) & 49.24(  2.02) \\
    QDA-WV & 22.75(  5.01) & 48.94(  2.18) & 48.98(  2.14) & 16.55(  4.87) & 45.82(  3.55) & 49.18(  2.11) \\
    SVM-MV & 18.39(  3.76) & 35.89(  4.53) & 38.74(  4.29) & 14.53(  3.81) & 32.04(  4.67) & 36.08(  4.12) \\
    DWD-MV & 17.17(  3.74) & 32.33(  4.07) & 37.69(  3.60) & 13.78(  3.30) & 26.93(  3.61) & 34.47(  4.11) \\
    MDEB-MV & 18.77(  5.01) & 32.73(  4.20) & 36.84(  4.01) & 14.71(  3.79) & 26.41(  4.04) & 32.58(  4.10) \\
    MDE-WV & 17.31(  5.24) & 31.33(  4.05) & 36.23(  4.05) & 12.95(  3.74) & 25.39(  3.92) & 31.39(  3.96) \\
    \hline
\multicolumn{7}{c}{Model (2) with $\rho = 0.5$}\\
 \hline
     PCF-LDA & 3.91(  4.23) & 1.66(  2.70) & 3.66(  2.87) & 0.49(  0.63) & 0.51(  0.62) & 1.58(  1.05) \\
    PCF-QDA & 6.77(  5.22) & 1.69(  2.78) & 3.67(  2.84) & 2.64(  2.46) & 0.49(  0.61) & 1.55(  1.09) \\
    PCF-SVM & 4.68(  4.55) & 1.66(  2.70) & 3.66(  2.87) & 0.88(  1.03) & 0.51(  0.62) & 1.58(  1.05) \\
    PCF-DWD & 6.16(  4.95) & 1.67(  2.74) & 3.62(  2.79) & 2.09(  1.65) & 0.50(  0.61) & 1.56(  1.05) \\
    PCF-MDEB & 5.71(  4.81) & 1.67(  2.74) & 3.64(  2.80) & 1.35(  1.28) & 0.51(  0.61) & 1.58(  1.05) \\
    PCF-YA & 10.15(  8.05) & 19.26( 10.01) & 26.92(  9.30) & 1.28(  2.02) & 18.82( 11.20) & 23.74(  9.72) \\
    LDA-MV & 40.53(  4.46) & 47.84(  3.49) & 46.83(  3.64) & 36.98(  4.04) & 46.53(  3.66) & 48.24(  3.69) \\
    LDA-WV & 39.52(  4.66) & 47.54(  3.30) & 46.68(  3.64) & 35.87(  4.31) & 46.18(  3.69) & 48.24(  3.46) \\
    QDA-MV & 3.44(  1.82) & 22.50(  3.79) & 21.30(  3.01) & 1.59(  0.94) & 31.51(  4.88) & 26.50(  3.95) \\
    QDA-WV & 1.30(  1.21) & 18.12(  3.61) & 17.01(  2.97) & 0.34(  0.38) & 28.94(  5.69) & 23.67(  3.75) \\
    SVM-MV & 40.25(  4.94) & 46.33(  3.49) & 45.69(  3.62) & 37.31(  3.77) & 47.06(  3.84) & 47.00(  3.31) \\
    DWD-MV & 37.27(  4.63) & 38.63(  4.15) & 39.86(  4.13) & 35.06(  3.73) & 36.31(  3.58) & 38.90(  3.37) \\
    MDEB-MV & 36.18(  6.17) & 42.69(  3.69) & 43.23(  3.98) & 32.50(  5.18) & 41.42(  4.01) & 43.39(  3.49) \\
    MDE-WV & 34.06(  7.11) & 41.88(  3.92) & 43.21(  3.71) & 29.82(  5.89) & 40.64(  4.26) & 43.45(  3.35) \\
    \hline
\multicolumn{7}{c}{Model (3) with $\rho = 0.5$}\\
 \hline
     PCF-LDA & 5.55(  4.46) & 17.37(  6.91) & 27.01(  7.95) & 2.50(  1.94) & 11.60(  5.58) & 20.40(  7.09) \\
    PCF-QDA & 6.15(  4.99) & 18.01(  7.26) & 27.33(  7.93) & 3.64(  3.10) & 13.87(  6.63) & 22.25(  7.74) \\
    PCF-SVM & 5.46(  4.14) & 17.35(  6.90) & 26.57(  6.52) & 2.79(  2.04) & 11.57(  5.45) & 20.46(  7.13) \\
    PCF-DWD & 5.96(  4.26) & 17.35(  6.71) & 27.17(  5.45) & 3.36(  2.48) & 11.36(  5.09) & 20.19(  6.96) \\
    PCF-MDEB & 7.38(  7.21) & 18.67(  6.47) & 29.07(  6.34) & 3.18(  2.31) & 11.62(  4.87) & 22.38(  6.07) \\
    PCF-YA & 11.48(  9.11) & 40.59(  8.34) & 44.36(  7.29) & 3.27(  3.00) & 39.45(  6.67) & 45.09(  5.57) \\
    LDA-MV & 37.96(  6.15) & 36.93(  4.64) & 36.82(  4.15) & 36.06(  5.55) & 36.54(  4.74) & 35.38(  3.49) \\
    LDA-WV & 37.28(  6.41) & 36.69(  4.49) & 36.58(  4.10) & 35.27(  6.13) & 36.29(  4.67) & 35.19(  3.54) \\
    QDA-MV & 26.55(  4.12) & 32.31(  3.32) & 33.99(  3.89) & 27.33(  6.55) & 33.33(  3.67) & 34.68(  3.40) \\
    QDA-WV & 24.04(  4.21) & 30.73(  3.26) & 33.03(  3.89) & 24.25(  6.29) & 31.58(  3.45) & 33.68(  3.66) \\
    SVM-MV & 37.96(  6.37) & 34.06(  4.57) & 34.87(  4.59) & 34.42(  6.13) & 33.76(  4.60) & 32.98(  3.81) \\
    DWD-MV & 30.16(  6.76) & 28.80(  4.00) & 31.65(  4.37) & 29.26(  6.13) & 24.68(  4.20) & 27.99(  3.67) \\
    MDEB-MV & 35.40( 11.35) & 31.05(  6.46) & 34.06(  5.79) & 37.01( 12.15) & 27.18(  6.03) & 30.54(  6.56) \\
    MDE-WV & 34.66( 11.99) & 30.52(  6.47) & 33.78(  5.83) & 36.53( 12.83) & 26.50(  6.16) & 30.16(  6.70) \\
    \hline
\multicolumn{7}{c}{Model (4) with $\rho = 0.5$}\\
 \hline
     PCF-LDA & 1.97(  1.70) & 34.18(  4.57) & 39.33(  4.51) & 0.15(  0.29) & 29.73(  3.58) & 33.32(  4.42) \\
    PCF-QDA & 6.81(  2.86) & 36.02(  4.76) & 41.06(  3.99) & 2.98(  1.54) & 33.95(  4.11) & 37.14(  4.02) \\
    PCF-SVM & 2.86(  2.10) & 34.21(  4.58) & 39.31(  4.51) & 0.48(  0.75) & 29.71(  3.55) & 33.34(  4.40) \\
    PCF-DWD & 4.46(  2.77) & 34.34(  5.06) & 39.89(  4.58) & 1.85(  1.72) & 29.56(  3.85) & 33.89(  4.88) \\
    PCF-MDEB & 3.76(  2.29) & 34.59(  5.40) & 40.09(  4.75) & 0.99(  1.04) & 29.51(  3.82) & 33.95(  5.01) \\
    PCF-YA & 5.08(  5.85) & 45.86(  5.66) & 47.98(  4.66) & 0.67(  1.77) & 43.20(  6.01) & 46.29(  4.63) \\
    LDA-MV & 39.29(  3.86) & 41.80(  3.51) & 44.54(  4.08) & 36.17(  4.39) & 38.78(  3.83) & 38.55(  4.62) \\
    LDA-WV & 38.01(  3.96) & 41.12(  4.05) & 44.44(  4.17) & 34.84(  4.42) & 37.92(  4.15) & 37.83(  4.73) \\
    QDA-MV & 3.25(  1.68) & 49.17(  1.90) & 49.28(  2.31) & 1.89(  1.18) & 45.65(  3.12) & 49.49(  1.76) \\
    QDA-WV & 0.55(  0.90) & 49.07(  2.10) & 49.22(  1.96) & 0.10(  0.23) & 45.32(  3.70) & 49.37(  1.61) \\
    SVM-MV & 38.46(  4.31) & 40.39(  3.67) & 40.08(  4.36) & 35.50(  4.34) & 40.23(  4.16) & 37.48(  4.40) \\
    DWD-MV & 34.99(  3.90) & 34.31(  3.90) & 38.57(  3.66) & 33.84(  4.01) & 30.94(  3.57) & 33.55(  4.15) \\
    MDEB-MV & 34.23(  5.55) & 35.98(  4.06) & 38.62(  4.03) & 30.99(  5.43) & 32.30(  3.39) & 32.71(  3.46) \\
    MDE-WV & 32.34(  6.18) & 34.92(  3.81) & 37.81(  4.07) & 28.63(  5.94) & 31.24(  4.01) & 31.80(  3.52) \\
     \hline
\end{tabular}
\end{footnotesize}
\end{table}

\end{document}